\documentclass[10pt,aps,preprintnumbers,prd,noshowpacs,nofootinbib,noshowkeys,floatfix,superscriptaddress]{revtex4-2}
\usepackage[english]{babel} 
\usepackage[dvips]{graphics,graphicx}
\usepackage[colorlinks=true,linktocpage=true,linkcolor=blue,citecolor=blue]{hyperref}
\usepackage[usenames,dvipsnames]{color}
\usepackage{amsmath, amssymb,oldgerm,amsfonts}
\usepackage{mathtools}
\usepackage{multirow}
\usepackage{enumitem} 		
\usepackage{longtable}
\usepackage{xcolor}
\usepackage{combelow}
\usepackage{dsfont}
\usepackage{cancel}
\usepackage[normalem]{ulem}  
\usepackage{slashed}
\usepackage[mathscr]{eucal}
\usepackage{mathtools}
\usepackage{ifthen}

\newcommand{\s}{y}

\newcommand{\C}{\mathcal{C}}

\renewcommand{\s}{\mathfrak{s}}
\newcommand{\f}{\mathfrak{f}}
\newcommand{\D}{\mathcal{D}}

\newcommand{\beq}{\begin{equation}}
\newcommand{\eeq}{\end{equation}}

\newcommand{\eqs}{Eqs.~}
\newcommand{\eq}{Eq.~}

\newcommand{\inmath}{\mathchar"3232}
\renewcommand{\in}{\mathrm{in}}
\newcommand{\out}{\mathrm{out}}
\newcommand{\momint}[3]{
\ifthenelse{\equal{#2}{}}
{\int\frac{\mathrm{d}^3\mathbf{#1}_{#3}}{(2\pi\hbar)^{3} 2 #1^0_{#3}}}
{\int\frac{\mathrm{d}^3\mathbf{#1}^{#2}_{#3}}{(2\pi\hbar)^{3#2} 2(#1^0_{#3})^{#2}}}
}
\newcommand{\avg}[1]{\left\langle#1\right\rangle}

\renewcommand{\k}{\mathbf{k}}

\renewcommand{\d}{\mathrm{d}}
\newcommand{\ket}[1]{\left|#1\right\rangle}
\newcommand{\bket}[1]{\bigg|#1\bigg\rangle}
\newcommand{\bra}[1]{\left\langle#1\right|}
\newcommand{\bbra}[1]{\bigg\langle#1\bigg|}
\newcommand{\braket}[2]{\left\langle#1\right|\left.#2\right\rangle}
\renewcommand{\Re}{\mathrm{Re}\,}
\renewcommand{\Im}{\mathrm{Im}\,}

\makeatletter
\DeclareFontFamily{OMX}{MnSymbolE}{}
\DeclareSymbolFont{MnLargeSymbols}{OMX}{MnSymbolE}{m}{n}
\SetSymbolFont{MnLargeSymbols}{bold}{OMX}{MnSymbolE}{b}{n}
\DeclareFontShape{OMX}{MnSymbolE}{m}{n}{
    <-6>  MnSymbolE5
   <6-7>  MnSymbolE6
   <7-8>  MnSymbolE7
   <8-9>  MnSymbolE8
   <9-10> MnSymbolE9
  <10-12> MnSymbolE10
  <12->   MnSymbolE12
}{}
\DeclareFontShape{OMX}{MnSymbolE}{b}{n}{
    <-6>  MnSymbolE-Bold5
   <6-7>  MnSymbolE-Bold6
   <7-8>  MnSymbolE-Bold7
   <8-9>  MnSymbolE-Bold8
   <9-10> MnSymbolE-Bold9
  <10-12> MnSymbolE-Bold10
  <12->   MnSymbolE-Bold12
}{}

\let\llangle\@undefined
\let\rrangle\@undefined
\DeclareMathDelimiter{\llangle}{\mathopen}%
                     {MnLargeSymbols}{'164}{MnLargeSymbols}{'164}
\DeclareMathDelimiter{\rrangle}{\mathclose}%
                     {MnLargeSymbols}{'171}{MnLargeSymbols}{'171}
\makeatother

\begin{document}

\title{Quantum kinetic theory with interactions for massive vector bosons}
\author{David Wagner}

\affiliation{Institute for Theoretical Physics, Goethe University,
Max-von-Laue-Str.\ 1, D-60438 Frankfurt am Main, Germany}
\affiliation{Department of Physics, West University of Timi\cb{s}oara, \\
Bd.~Vasile P\^arvan 4, Timi\cb{s}oara 300223, Romania}

\author{Nora Weickgenannt}

\affiliation{Institute for Theoretical Physics, Goethe University,
Max-von-Laue-Str.\ 1, D-60438 Frankfurt am Main, Germany}

\affiliation{Institut de Physique Th\'eorique, Universit\'e Paris Saclay, CEA, CNRS, F-91191 Gif-sur-Yvette, France}

\author{Enrico Speranza}

\affiliation{Illinois Center for Advanced Studies of the Universe \& Department of Physics, University of Illinois at Urbana-Champaign, Urbana, IL 61801, USA}



\begin{abstract}
We present a derivation of quantum kinetic theory for massive spin-1 particles from the Wigner-function formalism up to first order in an $\hbar$-expansion, including a general interaction term. Both local and nonlocal contributions are computed in a covariant fashion. It is shown that, up to first order in $\hbar$, the collision term takes the same form as in the case of spin-1/2 particles.
\end{abstract}



\maketitle

\section{Introduction}
The study of polarization phenomena in non-central heavy-ion collisions has become an active area of research in recent years \cite{Liang:2004ph,Voloshin:2004ha,Betz:2007kg,Becattini:2007sr,Becattini:2020ngo,Becattini:2022zvf}. The quark-qluon plasma (QGP) produced in such collisions behaves as a relativistic fluid, whose hydrodynamic gradients polarize the particles that are detected in experiments.
An example of such an effect is the global spin polarization of $\Lambda$-hyperons, which was measured at different collision energies by the STAR, ALICE and HADES collaborations~\cite{STAR:2017ckg,STAR:2018gyt,ALICE:2019onw,HADES:2022enx}. Despite global polarization data being well described by models which assume local equilibrium of spin degrees of freedom \cite{Becattini:2007sr,Becattini:2013vja,Becattini:2013fla,Becattini:2015ska,Becattini:2016gvu,Karpenko:2016jyx,Pang:2016igs,Xie:2017upb}, the polarization as a function of transverse momentum remains an active field of research \cite{Becattini:2020ngo,Liu:2021uhn,Fu:2021pok,Becattini:2021suc,Becattini:2021iol,Alzhrani:2022dpi}, spurring also the formulation of relativistic spin hydrodynamics  \cite{Montenegro:2017rbu,Florkowski:2017ruc,Florkowski:2017dyn,Florkowski:2018myy,Becattini:2018duy,Florkowski:2018fap,Montenegro:2018bcf,Weickgenannt:2019dks,Hattori:2019lfp,Montenegro:2020paq,Gallegos:2020otk,Weickgenannt:2020aaf,Garbiso:2020puw,Weickgenannt:2021cuo,Bhadury:2020puc,Gallegos:2021bzp,Shi:2020htn,Speranza:2020ilk,Bhadury:2020cop,Fukushima:2020ucl,Li:2020eon,Singh:2020rht,Bhadury:2021oat,Hongo:2021ona,Peng:2021ago,She:2021lhe,Wang:2021ngp,Wang:2021wqq,Sheng:2021kfc,Sheng:2022ssd,Hu:2021pwh,Hu:2022lpi,Singh:2022ltu,Daher:2022xon,Weickgenannt:2022zxs,Weickgenannt:2022jes,Gallegos:2022jow,Bhadury:2022qxd,Cao:2022aku,Biswas:2023qsw}.

In contrast, the theory behind explaining the spin alignment of vector mesons such as the $\phi$ and $K^{\star 0}$-mesons, which was measured by the ALICE and STAR collaborations~\cite{ALICE:2019aid,Mohanty:2021vbt,STAR:2022fan}, is not yet as developed. There have been notable steps taken towards providing a theoretical framework for this kind of polarization \cite{Liang:2004xn,Yang:2017sdk,Sheng:2019kmk,Sheng:2020ghv,Xia:2020tyd,Goncalves:2021ziy,Muller:2021hpe,Sheng:2022wsy,Sheng:2022ffb,Wagner:2022gza,Li:2022vmb,Kumar:2023ghs}. 
In particular, in Ref.~\cite{Wagner:2022gza} we discussed a new mechanism which relates the spin alignment to the shear stress of the fluid and provided a formula which can be used for phenomenological applications. In this paper, we  systematically formulate quantum kinetic theory for massive spin-1 particles based on the Wigner-function formalism used in Ref.~\cite{Wagner:2022gza}.

Vector mesons have a richer structure as compared to spin-1/2 particles. While the spin-density matrix of spin-1/2 particles only has three independent entries which determine the polarization vector, spin-1 particles feature eight internal degrees of freedom~\cite{Leader:2001}. 
Three of these entries are again associated to a polarization vector, while the remaining five quantify the so-called tensor polarization. In present heavy-ion collision experiments, since  strong decays of vector particles are studied, only the effects of tensor polarization can be accessed through the $00$-element of the spin-density matrix. Thus, the spin alignment of vector mesons can be considered as a genuine spin-1 effect, which necessitates the development of an adequate formalism.

As opposed to approaches put forward in Refs.~\cite{Liang:2004xn,Yang:2017sdk,Sheng:2019kmk,Sheng:2020ghv,Xia:2020tyd,Muller:2021hpe,Sheng:2022wsy,Sheng:2022ffb} where the properties of the vector mesons are determined through the polarization of the individual quarks, the aim of this work is to construct a quantum kinetic theory using the Wigner-function formalism where the effective degrees of freedom are massive spin-1 fields. Employing a semiclassical expansion, the equations of motion are truncated at first order in the Planck constant, resulting in a Boltzmann-type equation.
This procedure stands in line with Ref. \cite{Weickgenannt:2021cuo}, where such a formalism, accounting for general interactions, was developed for spin-1/2 particles, and later used as a foundation for the formulation of spin-hydrodynamics \cite{Weickgenannt:2022zxs,Weickgenannt:2022qvh}.

This paper is structured as follows. In Sec. \ref{sec:kin_theory}, we discuss the quantum kinetic theory for massive spin-1 particles and obtain a Boltzmann equation for the Wigner function. 
In Sec.~\ref{sec:coll}, we expand the collision integral on the right-hand side of the Boltzmann equation in terms of the Wigner function, restricting ourselves to binary elastic collisions. 
Section \ref{sec:eq} focuses on the derivation of the local-equilibrium distribution function that makes the previously obtained collision term vanish, while Sec. \ref{sec:spin_hydro} discusses the implications for the formulation of spin-1 hydrodynamics. Finally, in Sec. \ref{sec:conclusion} we give the conclusions and outlook.

We use the following notation and conventions: $a\cdot b\coloneqq  a^\mu b_\mu$,
$a_{[\mu}b_{\nu]}\coloneqq  a_\mu b_\nu-a_\nu b_\mu$, $a_{(\mu}b_{\nu)}\coloneqq  a_\mu b_\nu+a_\nu b_\mu$, $g_{\mu \nu} \coloneqq  \mathrm{diag}(1,-1,-1,-1)$,
$\epsilon^{0123} = - \epsilon_{0123} \coloneqq  1$, and repeated indices are summed over. Projection operators parallel and orthogonal to the four-momentum $k^\mu$ are denoted by $E^{\mu\nu}\coloneqq  k^\mu k^\nu /k^2$ and $K^{\mu\nu}\coloneqq  g^{\mu\nu}-E^{\mu\nu}$, respectively.
We use natural units with $c=k_B=1$, but do not set $\hbar$ to unity in order to allow for a clear semiclassical expansion.

\section{Kinetic theory for interacting vector bosons}
\label{sec:kin_theory}
We start from the Lagrangian for a charged vector field $\hat{V}^\mu$,
\begin{equation}
\label{Lagrangian}
\hat{\mathcal{L}}=-\hbar  \left(\frac12 \hat{V}^{\dagger\mu\nu} \hat{V}_{\mu\nu} -\frac{m^2}{\hbar^2}\hat{V}^{\dagger\mu}\hat{V}_\mu\right)+\hat{\mathcal{L}}_{\text{int}}\;,
\end{equation}
where $\hat{V}^{\mu\nu}\coloneqq \partial^{[\mu} \hat{V}^{\nu]}$ and $\hat{\mathcal{L}}_{\text{int}}$ is a general interaction Lagrangian, which we assume not to depend on the derivatives of the field. Defining 
\begin{equation}
\label{source}
\hat{\rho}^\nu \coloneqq -\frac{1}{\hbar }\frac{\partial \hat{\mathcal{L}}_{\text{int}}}{\partial \hat{V}^\dagger_\nu} \;,
\end{equation}
the equations of motion read
\begin{equation}
\left(\Box +\frac{m^2}{\hbar^2}    \right)\hat{V}^\nu-\partial^\nu \partial\cdot \hat{V} =\hat{\rho}^\nu\;,
\label{EoM}
\end{equation}
from which the constraint equation
\begin{equation}
\label{constraint}
\partial \cdot \hat{V}= \frac{\hbar^2}{m^2} \partial\cdot \hat{\rho}
\end{equation}
follows. 

The Wigner function is defined as \cite{Vasak:1987um, Elze:1986hq,Elze:1989un,Huang:2020kik,Hattori:2020gqh,Weickgenannt:2022jes}
\begin{equation}
\label{Wigner_function}
W^{\mu\nu}\coloneqq -\frac{2}{(2\pi\hbar)^4\hbar} \int \d^4 v e^{-ik\cdot v/\hbar} \avg{:\hat{V}_+^{\dagger\mu} \hat{V}_-^\nu:}\;,
\end{equation}
where $\hat{V}_\pm^\mu \coloneqq \hat{V}^\mu (x\pm v/2)$, and its evolution equations are found via the Bopp operators $D^\mu\coloneqq  k^\mu +\frac{i\hbar}{2}\partial^\mu$ \cite{Bopp1956}, which fulfill
\begin{subequations}\label{Ds}
\begin{eqnarray}
\label{D}
D^\mu W^{\alpha\beta}(x,k)&=&-i\hbar \frac{2}{(2\pi\hbar)^4\hbar} \int \d^4 v e^{-ik\cdot v/\hbar} \avg{:\hat{V}^{\dagger\alpha}_+ \partial^\mu \hat{V}^\beta_-:}\;,\\
D^{*\mu} W^{\alpha\beta}(x,k)&=&i\hbar \frac{2}{(2\pi\hbar)^4\hbar} \int \d^4 v e^{-ik\cdot v/\hbar} \avg{:\left(\partial^\mu \hat{V}^{\dagger\alpha}_+\right)  \hat{V}^\beta_-:}\;.
\label{Dstar}
\end{eqnarray}
\end{subequations}
Acting with appropriate combinations of the Bopp operators to make use of the field equations \eqref{EoM} and defining the collision term
\begin{equation}
\label{eq:coll_int_1}
C^{\mu\nu}\coloneqq -\frac{2}{(2\pi\hbar)^4} \int \d^4 y\, e^{-ik\cdot y/\hbar} \avg{:\hat{V}^{\dagger\mu}_+\hat{\rho}^\nu_- :}\; ,
\end{equation}
we obtain the equation of motion for the Wigner function, 
\begin{equation}
\left( D^2+m^2   \right)W^{\mu\nu} - \frac{\hbar}{m^2}D^\nu D_\alpha C^{\mu\alpha}=-\hbar C^{\mu\nu}\;.
\label{eom567}
\end{equation}
In addition to the fact that the Wigner function is hermitian, we employed the constraint equation \eqref{constraint}, whose analogue for the Wigner function reads
\begin{equation}
D_\mu W^{\nu\mu}=\frac{\hbar}{m^2} D_\mu C^{\nu\mu}\;.\label{cons_Bopp}
\end{equation}

In order to disentangle the equations of motion for the Wigner function into several expressions determining its time evolution as well as its structure in momentum space, we define two hermitian combinations of the collision term \eqref{eq:coll_int_1},
\begin{subequations}\label{C_M_def}
\begin{eqnarray}
\mathcal{C}^{\mu\nu}&\coloneqq &-\frac{i}{(2\pi\hbar)^4} \int \d^4 v e^{-ik\cdot v/\hbar} \avg{:V_+^{\dagger\mu}\rho_-^\nu -\rho_+^{\dagger\mu}V_-^\nu:}\equiv \frac{i}{2}\left( C^{\mu\nu} -C^{*\nu\mu}   \right)\;,
\label{def_C}\\
\delta M^{\mu\nu}&\coloneqq &\frac{1}{(2\pi\hbar)^4} \int \d^4 v e^{-ik\cdot v/\hbar} \avg{:V_+^{\dagger\mu}\rho_-^\nu +\rho_+^{\dagger\mu}V_-^\nu:}\equiv -\frac{1}{2}\left( C^{\mu\nu}  +C^{*\nu\mu}   \right)\;.
\label{def_M}
\end{eqnarray}
\end{subequations}
From the constraint equation \eqref{cons_Bopp} and its complex conjugate, we obtain
\begin{subequations}\label{constr2}
\begin{eqnarray}
k_\mu W_S^{\mu\nu}-\frac{i\hbar}{2}\partial_\mu W_A^{\mu\nu} &=& \frac{\hbar}{m^2}\left[k_\mu \left(i\C^{\mu\nu}_A-\delta M^{\mu\nu}_S\right)+\frac{\hbar}{2}\partial_\mu\left(\C^{\mu\nu}_S+i\delta M^{\mu\nu}_A\right)   \right]\;,\\
k_\mu W_A^{\mu\nu}-\frac{i\hbar}{2}\partial_\mu W_S^{\mu\nu}&=&\frac{\hbar}{m^2}\left[k_\mu \left(i\C^{\mu\nu}_S-\delta M^{\mu\nu}_A\right)+\frac{\hbar}{2}\partial_\mu\left(\C^{\mu\nu}_A+i\delta M^{\mu\nu}_S\right)   \right]\;,
\end{eqnarray}
\end{subequations}
where we split both $W^{\mu\nu}$ and the collision terms $\delta M^{\mu\nu},\C^{\mu\nu}$ into symmetric and antisymmetric parts , denoted with the subscript $S$ and $A$, respectively. Note that the symmetric parts of $\delta M^{\mu\nu}, W^{\mu\nu}$ and $\C^{\mu\nu}$ are real, while their antisymmetric parts are imaginary.
Using this fact, we obtain the Boltzmann-like equations for the symmetric and antisymmetric parts of the Wigner function from \eq\eqref{eom567},
\begin{subequations}\label{Boltzmann_SA}
\begin{eqnarray}
k\cdot \partial W_S^{\mu\nu}&=&\C^{\mu\nu}_S-\frac{1}{2m^2}\left[ \left(k_\alpha k^{(\mu}-\frac{\hbar^2}{4}\partial_\alpha \partial^{(\mu}\right)\left(\C^{\nu)\alpha}_S-i\delta M_A^{\nu)\alpha}\right) +\frac{\hbar}{2}\left(k_\alpha \partial^{(\mu}+\partial_\alpha k^{(\mu}\right)\left(i\C_A^{\nu)\alpha}+\delta M^{\nu)\alpha}_S\right)  \right]\label{Boltzmann_S} \;,\qquad\\
k\cdot \partial W_A^{\mu\nu}&=&\C^{\mu\nu}_A-\frac{1}{2m^2}\left[\left(k_\alpha k^{[\mu}-\frac{\hbar^2}{4}\partial_\alpha \partial^{[\mu}\right)\left(i\delta M_S^{\nu]\alpha}-\C^{\nu]\alpha}_A\right) -\frac{\hbar}{2}\left(k_\alpha \partial^{[\mu}+\partial_\alpha k^{[\mu}\right)\left(i\C_S^{\nu]\alpha}+\delta M^{\nu]\alpha}_A\right)  \right]\label{Boltzmann_A}\;.
\end{eqnarray}
\end{subequations}

In order to further clarify the structure of the constraint and evolution equations \eqref{constr2} and \eqref{Boltzmann_SA}, it is helpful to decompose the Wigner function and all related quantities with respect to the four-momentum $k^\mu$,
\begin{subequations}\label{decomp_1}
\begin{eqnarray}
W_S^{\mu\nu}&=&E^{\mu\nu} f_E + \frac{k^{(\mu}}{2k}F_S^{\nu)}+F_K^{\mu\nu}+K^{\mu\nu}f_K\;,\quad
W_A^{\mu\nu}=i\frac{k^{[\mu}}{2k}F_A^{\nu]}+i\epsilon^{\mu\nu\alpha\beta} \frac{k_\alpha}{m}G_\beta\;,\\
\C^{\mu\nu}_S&=&E^{\mu\nu} \C_E + \frac{k^{(\mu}}{2}\C_S^{\nu)}+\C_K^{\mu\nu}+K^{\mu\nu}\C_K\;,\quad
\C^{\mu\nu}_A=i\frac{k^{[\mu}}{2k}\C_A^{\nu]}+i\epsilon^{\mu\nu\alpha\beta} \frac{k_\alpha}{m}\C_{G,\beta}\;,\\
\delta M^{\mu\nu}_S&=&E^{\mu\nu} \D_E + \frac{k^{(\mu}}{2k}\D_S^{\nu)}+\D_K^{\mu\nu}+K^{\mu\nu}\D_K\;,\quad
\delta M^{\mu\nu}_A=i\frac{k^{[\mu}}{2k}\D_A^{\nu]}+i\epsilon^{\mu\nu\alpha\beta} \frac{k_\alpha}{m}\D_{G,\beta}\;,
\end{eqnarray}
\end{subequations}
where $F_S\cdot k=F_A\cdot k=G\cdot k=0$, $F_K^{\mu\nu}k_\nu=0$, and $F_K^{\mu\nu}$ is symmetric and traceless. Analogous properties hold for the components of the collision terms $\C^{\mu\nu}$, $\delta M^{\mu\nu}$. We remind the reader that the projection operators with respect to the four-momentum used above are defined as $E^{\mu\nu}\coloneqq k^\mu k^\nu /k^2$ and $K^{\mu\nu}\coloneqq g^{\mu\nu}-E^{\mu\nu}$.

In the remainder of this section, we aim at performing a semiclassical expansion to next-to-leading order in powers of $\hbar$, i.e., setting $B\coloneqq \sum_{j=0}^\infty \hbar^j B^{(j)}$ for any quantity $B$ in Eqs. \eqref{constr2},  \eqref{Boltzmann_SA} and truncating the sum at $j=1$. We note that, since factors of $\hbar$ are always accompanied by a gradient of the Wigner function, this expansion is effectively a gradient expansion.
At this point it is evident that, in the course of such an expansion in $\hbar$, the effect of the constraint equations \eqref{constr2} consists of expressing $f_E$, $F_S^\mu$, and $F_A^\mu$ (i.e., the components of the Wigner function which are parallel to the four-momentum in at least one index) in terms of $f_K$, $F_K^{\mu\nu}$, and $G^\mu$, while the kinetic equations \eqref{Boltzmann_SA} determine the time evolution of the latter quantities.

In order to expand the kinetic equations to first order in the Planck constant, we have to clarify which parts of the collision terms enter at leading order. 
Using the definition of the collision term \eqref{eq:coll_int_1} and the constraint \eqref{constraint}, we obtain
\begin{equation}
   \left(k_\mu -\frac{i\hbar}{2}\partial_\mu\right) C^{\mu\nu}=\mathcal{O}(\hbar)\; ,\qquad \left(k_\mu +\frac{i\hbar}{2}\partial_\mu\right) C^{*\mu\nu}=\mathcal{O}(\hbar)\; ,
\end{equation}
from which it follows that $\C_E^{(0)}=\D_E^{(0)}=0$. Note that there are no such constraints on the other components of $\mathcal{C}^{\mu\nu}$ and $\delta M^{\mu\nu}$, which can in principle enter at zeroth order already. However, following Refs. \cite{Weickgenannt:2020aaf, Weickgenannt:2022jes}, we consider a situation where no initial large (vector- or tensor-) polarization is present. In this case we conclude that $G^{(0)\mu}=0$ and $F_K^{(0)\mu\nu}=0$ as well as $\C^{(0)\mu}_S=\C^{(0)\mu}_A=\D^{(0)\mu}_S=\D^{(0)\mu}_A=0$, which follows from the fact that there are no vector or tensor structures at our disposal at order $\mathcal{O}(1)$ which possess the required symmetries of the aforementioned terms.

With these simplifications, we obtain from the real parts of \eqs\eqref{constr2}
\begin{subequations}\label{constr_coll}
\begin{eqnarray}
f_E&=&\frac{\hbar^2}{4k^2} K^{\alpha\beta} \partial_\alpha \partial_\beta f_K^{(0)}-\frac{\hbar}{m^2}\D_E+\mathcal{O}(\hbar^3)\;,\\
F_S^{\nu}&=&\mathcal{O}(\hbar^2)\;,\\
k F_A^{\nu}&=&\hbar K^{\nu\mu}\partial_\mu f_K^{(0)}+\mathcal{O}(\hbar^2)\;.
\label{81c}
\end{eqnarray}
\end{subequations}
Furthermore, from \eqs\eqref{Boltzmann_SA} we obtain a simple form of the Boltzmann-like equations for the independent parts of the Wigner function,
\begin{subequations}
\label{kin_eqs}
\begin{eqnarray}
k\cdot \partial f_K&=&\C_K+\mathcal{O}(\hbar^2)\; , \label{Boltzmann_scalar_1}\\
k\cdot \partial F_K^{\mu\nu}&=& \C^{\mu\nu}_K+\mathcal{O}(\hbar^2)\; ,\label{Boltzmann_tensor_1}\\
k\cdot \partial G^{\mu}&=& \C_G^{\mu}+\mathcal{O}(\hbar^2)\label{Boltzmann_vector_1}\;,
\end{eqnarray}
\end{subequations}
while the mass-shell equations follow from the real part of Eq. \eqref{eom567},
\begin{subequations}
\label{massshell_eqs}
\begin{eqnarray}
(k^2-m^2) f_K&=&\hbar\D_K+\mathcal{O}(\hbar^2)\; , \label{massshell_scalar_1}\\
(k^2-m^2) F_K^{\mu\nu}&=& \hbar\D^{\mu\nu}_K+\mathcal{O}(\hbar^2)\; ,\label{massshell_tensor_1}\\
(k^2-m^2)G^{\mu}&=& \hbar\D_G^{\mu}+\mathcal{O}(\hbar^2)\label{massshell_vector_1}\;.
\end{eqnarray}
\end{subequations}

In order to account for the degrees of freedom of the Wigner function connected to spin, in analogy with Refs. \cite{Weickgenannt:2020aaf,Weickgenannt:2021cuo,Weickgenannt:2022jes}, we may enlarge the phase space by introducing an additional variable $\s^\mu$, together with a respective measure
\begin{equation}
\d S(k)\coloneqq \frac{3m}{2\sigma \pi}\d^4 \s \delta(\s^2+\sigma^2)\delta(k\cdot \s)\;, \quad \sigma^2 \coloneqq 2\;,\label{dS}
\end{equation}
such that
\begin{eqnarray}
\int \d S (k)=3\;,\quad \int \d S(k) \s^\mu \s^\nu =-2K^{\mu\nu}\;,\quad \int \d S(k) K^{\mu\nu}_{\rho\sigma}\s^{\rho} \s^{\sigma} \s^\alpha \s^\beta =\frac{8}{5}K^{\mu\nu,\alpha\beta},
\end{eqnarray}
where $K^{\mu\nu}_{\alpha\beta}\coloneqq \frac12 K^\mu_{(\alpha} K^\nu_{\beta)}-\frac13 K^{\mu\nu}K_{\alpha\beta}$,
and
\begin{equation}
\int \d S(k) \s^{\mu_1} \cdots \s^{\mu_{2n+1}} =0 \quad \text{for} \;\; n \inmath  \mathbb{N}\;.
\end{equation}
Note that the measure \eqref{dS} differs from the spin-1/2 case \cite{Weickgenannt:2021cuo,Weickgenannt:2020aaf,Speranza:2020ilk} in the normalization of the spin vector $\s^\mu$, whose squared length is given by $\sigma^2=2\equiv s(s+1)$ with $s=1$, while the normalization of the phase-space volume is given by the spin degeneracy, as expected. We remark that the introduction of the continuous variable $\s^\mu$ does not imply a classical treatment of spin. Instead, it provides a tool to define a scalar distribution function that still contains all information related to spin degrees of freedom, which are contained in $G^\mu$ and $F_K^{\mu\nu}$.

Defining a distribution function in this enlarged phase space
\begin{equation}
\f(x,k,\s)\coloneqq  f_K-\s\cdot G+\frac58\s^{\mu}\s^{\nu} F_{K,\mu\nu}   \;,
\end{equation}
which fulfills
\begin{eqnarray}
\frac{1}{3}\int \d S(k) \f=f_K\;,\quad \frac{1}{2}\int \d S(k) \s^\mu \f  =G^\mu \;,\quad \int \d S(k) K^{\mu\nu}_{\alpha\beta}\s^{\alpha} \s^{\beta} \f =F_K^{\mu\nu}\;,
\end{eqnarray}
Eqs. \eqref{kin_eqs} and \eqref{massshell_eqs} become
\begin{subequations}\label{kin_eqs_new}
\begin{eqnarray}
k\cdot \partial \f &=&\mathfrak{C}[\f]\;,\label{Boltzmann_f}\\
(k^2-m^2)\f&=&\hbar\mathfrak{M}\;,\label{mass_shell}
\end{eqnarray}
\end{subequations}
where we defined
\begin{equation}\label{C_M_def_new}
\mathfrak{C}[\f]\coloneqq \C_K -\s\cdot \C_G +\frac58\s_\mu \s_\nu \C_K^{\mu\nu} \;,\qquad
\mathfrak{M}\coloneqq \D_K-\s\cdot \D_{G}+\frac58\s_\mu \s_\nu \D_{K}^{\mu\nu}\;.
\end{equation}
In analogy with the decomposition of the Wigner function, the components of $\mathcal{C}^{\mu\nu}$ and $\delta M^{\mu\nu}$ are defined as
\begin{subequations}
\begin{eqnarray}
\C_K&\coloneqq & \frac13 K_{\mu\nu} \C^{\mu\nu} \;,\quad \C_G^\mu\coloneqq -\frac{i}{2}\epsilon^{\mu\nu\alpha\beta} \frac{k_\nu}{m} \C_{\alpha\beta} \;,\quad \C_K^{\mu\nu}\coloneqq   K^{\mu\nu}_{\alpha\beta} \C^{\alpha\beta}\;,\\
\D_K&\coloneqq  & \frac13 K_{\mu\nu} \delta M^{\mu\nu} \;,\quad \D_G^\mu\coloneqq -\frac{i}{2}\epsilon^{\mu\nu\alpha\beta} \frac{k_\nu}{m} \delta M_{\alpha\beta} \;, \quad\D_K^{\mu\nu}\coloneqq  K^{\mu\nu}_{\alpha\beta} \delta M^{\alpha\beta}\;.
\end{eqnarray}
\end{subequations}
Similar to the spin-1/2-case discussed in Ref. \cite{Weickgenannt:2021cuo}, the solution of the mass-shell equation \eqref{mass_shell} is given by
\begin{equation}
\f(x,k,\s)=\delta(k^2-M^2)f(x,k,\s)\;,
\end{equation}
where $M^2\coloneqq  m^2+\hbar\delta m^2(x,k,\s)$, and
\begin{equation}
\hbar \delta(k^2-m^2)\delta m^2(x,k,\s) f(x,k,\s)=\hbar\mathfrak{M}(x,k,\s)+\mathcal{O}(\hbar^2)\;.
\end{equation}
We will show in Appendix \ref{app:offshell} that the off-shell effects cancel in the Boltzmann equation.

\section{Evaluating the collision term}
\label{sec:coll}

In this section we follow the approach by De Groot et al. \cite{DeGroot:1980dk} to expand the collision operator appearing in Eq. \eqref{Boltzmann_f} in terms of the distribution function $f$. The steps performed here are similar to the ones for scalar particles discussed in \cite{DeGroot:1980dk}, but differ in some aspects due to the nontrivial spin structure of vector particles. For this reason we recap some of the steps of Ref. \cite{DeGroot:1980dk} here, with more technical details explained in Appendix \ref{app:rewriting}. 

The main idea is to express any operator in terms of the asymptotic initial states (``in''-states), which are defined by
\begin{equation}
\ket{k^n;\lambda^n}_\in \coloneqq  \hat{a}^\dagger_\in (k^n,\lambda^n) \ket{0}\;,
\end{equation}
where
\begin{equation}
k^n\coloneqq k_1^\mu ,k_2^\mu , \cdots , k_n^\mu\;,\quad \lambda^n\coloneqq \lambda_1,\lambda_2,\cdots,\lambda_n \;,\quad \hat{a}^\dagger_\in (k^n,\lambda^n)\coloneqq \hat{a}^\dagger_\in (k_1,\lambda_1)\hat{a}^\dagger_\in (k_2,\lambda_2)\cdots \hat{a}^\dagger_\in (k_n,\lambda_n)\;.
\end{equation}
These states form a complete and orthogonal basis of the Fock space, i.e., we have the relations
\begin{subequations}
\begin{eqnarray}
\prescript{}{\in}{\braket{k;\lambda}{k';\lambda'}}_{\in}&=& (2\pi \hbar)^3 2 k^0 \delta^{(3)}(\mathbf{k}-\mathbf{k}')\delta_{\lambda\lambda'}\;,\label{orthogonality}\\
\label{completeness}
\mathds{1}&=&\sum_{n=0}^\infty \frac{1}{n!} \sum_{\lambda^n} \momint{k}{n}{} \ket{k^n;\lambda^n}_{\in}\prescript{}{\in}{\bra{k^n;\lambda^n}} \;.
\end{eqnarray}
\end{subequations}
Here we defined
\begin{equation}
\momint{k}{n}{}\coloneqq \momint{k}{}{1}\momint{k}{}{2}\cdots \momint{k}{}{n}\;,\quad 
\sum_{\lambda^n}\coloneqq \sum_{\lambda_1=1}^3 \sum_{\lambda_2=1}^3\cdots \sum_{\lambda_n=1}^3\;.
\end{equation}
The factorial in Eq. \eqref{completeness} is needed to account for double counting, such that the Fock space is spanned by all distinct ``in''-states. Note that the same completeness relation also holds for the  asymptotic final states (``out''-states). 
Using these creation and annihilation operators, we define the ``in''-fields
\begin{equation}
\hat{V}^\mu_\in (x)\coloneqq  \sqrt{\hbar} \sum_\sigma\int \frac{\d^4 k}{(2\pi\hbar)^3} \Theta(k^0) \delta(k^2-m^2) e^{-\frac{i}{\hbar}k\cdot x } \epsilon^{(\sigma)\mu}(k) \hat{a}_\in (k,\sigma)\;,\label{V_in}
\end{equation}
where the prefactor is needed to recover the correct units of the vector field. Here, $\epsilon^{(\sigma)\mu}(k)$ are polarization vectors, which fulfill the following orthogonality and completeness relations
\begin{subequations}
\begin{eqnarray}
\epsilon^{*(\lambda)\mu}(k)\epsilon^{(\lambda')}_\mu (k) &=&-\delta_{\lambda\lambda'}\;, \\
\sum_{\lambda} \epsilon^{*(\lambda)\mu}(k)\epsilon^{(\lambda)\nu}(k)&=&-K^{\mu\nu}\;.
\end{eqnarray}
\end{subequations}
Using Eq. \eqref{V_in}, we define the ``in''-Wigner function
\begin{equation}
W_\in^{\mu\nu}(x,k)\coloneqq -\frac{2}{(2\pi\hbar)^4\hbar} \int \d^4 v e^{-\frac{i}{\hbar} k\cdot v} \avg{\hat{V}^{\dagger\mu}_\in \left(x+\frac{v}{2}\right) \hat{V}^\nu_\in \left(x-\frac{v}{2}\right)   }\;.\label{W_in_def}
\end{equation}
In a subsequent step, we use the fact that an arbitrary operator may be expressed in terms of the ``in''-Wigner functions. 
This step contains the assumption of molecular chaos, thus rendering the evolution of the Wigner function irreversible, cf. Appendix \ref{app:rewriting} and Ref. \cite{DeGroot:1980dk}.
Then, one can use this expression to express both the Wigner function itself and the collision kernel $\mathcal{C}^{\mu\nu}$ in terms of $W_\in^{\mu\nu}$.
Both calculations are detailed in Appendix \ref{app:scattering}. 

The Wigner function itself can simply be written as $W^{\mu\nu}=W_\in^{\mu\nu}+\cdots\,$, where the dots indicate contributions of higher order in the density, which we neglect in this work in the collision term \cite{DeGroot:1980dk, Weickgenannt:2021cuo}. We show in Appendix \ref{app:scattering} that the components of the collision term that are orthogonal to the four-momentum (marked by a subscript ``$\perp$'') read to lowest order in the density and to first order in $\hbar$
\begin{eqnarray}
\mathcal{C}^{\mu\nu}_{\perp,\text{on-shell}} &=& \frac{(2\pi\hbar)^7}{4} \int \frac{\d^3 \mathbf{k}_1}{(2\pi\hbar)^32k_1^0}\int \frac{\d^3 \mathbf{k}_2}{(2\pi\hbar)^32k_2^0}\int \frac{\d^3 \mathbf{k}'}{(2\pi\hbar)^32k'^0}  M^{\gamma_1\gamma_2 \delta_1\delta_2}M^{\zeta_1\zeta_2 \eta_1\eta_2} \int \d^4 u^2\nonumber\\
&&\times K^\mu_{\mu'}K^\nu_{\nu'} \left(K-\frac{U_1+U_2}{2}\right)^{\mu'\alpha} \left(K+\frac{U_1+U_2}{2}\right)^{\nu'\beta} \Bigg( \delta^{(4)}(k+k'-k_1-k_2) g_{\alpha\delta_1}g_{\beta\zeta_1}K'_{\delta_2\zeta_2} \nonumber\\ 
&&\times  \left\{\delta^{(4)}(u_1)W_{\text{on-shell}}^{\alpha_1\beta_1}(x,k_1)-i\hbar \left[\partial^\rho_{u_1}\delta^{(4)}(u_1)\right]\partial_\rho W_{\text{on-shell}}^{\alpha_1\beta_1}(x,k_1)\right\}\left(K_1-\frac{U_1}{2}\right)_{\gamma_1\alpha_1}\left(K_1+\frac{U_1}{2}\right)_{\beta_1\eta_1}\nonumber\\
&&\times\left\{\delta^{(4)}(u_2)W_{\text{on-shell}}^{\alpha_2\beta_2}(x,k_2)-i\hbar \left[\partial^\rho_{u_2}\delta^{(4)}(u_2)\right]\partial_\rho W_{\text{on-shell}}^{\alpha_2\beta_2}(x,k_2)\right\}\left(K_2-\frac{U_2}{2}\right)_{\gamma_2\alpha_2}\left(K_2+\frac{U_2}{2}\right)_{\beta_2\eta_2}\nonumber\\
&-& \frac12 \left\{\delta^{(4)}(u_1)W_{\text{on-shell}}^{\alpha_1\beta_1}\left(x,k-\frac{u_2}{2}\right)-i\hbar \left[\partial_{u_1}^\rho \delta^{(4)}(u_1)\right]\partial_\rho W_{\text{on-shell}}^{\alpha_1\beta_1}\left(x,k-\frac{u_2}{2}\right)\right\}\nonumber\\
&&\times \left\{\delta^{(4)}(u_2)W_{\text{on-shell}}^{\alpha_2\beta_2}(x,k')-i\hbar\left[\partial_{u_2}^\rho \delta^{(4)}(u_2)\right]\partial_\rho W_{\text{on-shell}}^{\alpha_2\beta_2}\left(x,k'\right) \right\}   \left(K'-\frac{U_2}{2}\right)_{\gamma_2\alpha_2}K_{1,\delta_1\zeta_1}K_{2,\delta_2\zeta_2}\nonumber\\
&&\times g_{\alpha\alpha_1}g_{\beta\gamma_1}
\left(K+\frac{U_1-U_2}{2}\right)_{\beta_1\eta_1}\left(K'+\frac{U_2}{2}\right)_{\beta_2\eta_2}\delta^{(4)}\left(k+k'-k_1-k_2+\frac{u_1}{2}\right)\nonumber\\
&-& \frac12
\left\{\delta^{(4)}(u_1)W_{\text{on-shell}}^{\alpha_1\beta_1}\left(x,k+\frac{u_2}{2}\right)-i\hbar\left[\partial_{u_1}^\rho \delta^{(4)}(u_1)\right]\partial_\rho W_{\text{on-shell}}^{\alpha_1\beta_1}\left(x,k+\frac{u_2}{2}\right) \right\}\nonumber\\
&&\times \left\{\delta^{(4)}(u_2)W_{\text{on-shell}}^{\alpha_2\beta_2}(x,k')-i\hbar\left[\partial_{u_2}^\rho \delta^{(4)}(u_2)\right]\partial_\rho W_{\text{on-shell}}^{\alpha_2\beta_2}\left(x,k'\right) \right\}   \left(K'-\frac{U_2}{2}\right)_{\gamma_2\alpha_2}K_{1,\delta_1\zeta_1}K_{2,\delta_2\zeta_2}\nonumber\\
&&\times g_{\alpha\eta_1}g_{\beta\beta_1}
\left(K+\frac{U_2-U_1}{2}\right)_{\gamma_1\alpha_1}\left(K'+\frac{U_2}{2}\right)_{\beta_2\eta_2}\delta^{(4)}\left(k+k'-k_1-k_2-\frac{u_1}{2}\right)\Bigg)\;.\label{eq:coll_int_full}
\end{eqnarray}
In this equation, the fourth-rank tensors $M$ are the tree-level vertices of the theory, defined in Eq. \eqref{eq:t_to_M}.
Compared to Ref. \cite{Weickgenannt:2021cuo}, we neglected the contribution of collisions that exchange only spin, but not momentum, since this term can be considered as a modification of the left-hand side of the Boltzmann equation, cf. Ref. \cite{Wagner:2022amr}.
In the following subsections, we evaluate the local and nonlocal components of Eq. \eqref{eq:coll_int_full}.

\subsection{Local collisions}
The local collision term is obtained by taking into account only the contributions in Eq. \eqref{eq:coll_int_full} that are proportional to $\delta^{(4)}(u_1)\delta^{(4)}(u_2)$, and it is thus given by
\begin{eqnarray}
    \mathcal{C}^{\text{local},\,\mu\nu}_{\perp,\text{on-shell}}(x,k) &=& \frac{(2\pi\hbar)^7}{4}  \int \frac{\d^3 \mathbf{k}_1}{(2\pi\hbar)^32k_1^0}\int \frac{\d^3 \mathbf{k}_2}{(2\pi\hbar)^32k_2^0}\int \frac{\d^3 \mathbf{k}'}{(2\pi\hbar)^32k'^0} \delta^{(4)}(k+k'-k_1-k_2) M^{\gamma_1\gamma_2 \delta_1\delta_2}M^{\zeta_1\zeta_2 \eta_1\eta_2}\nonumber\\
&\times&  \bigg[  W_{\text{on-shell}}^{\alpha_1\beta_1}(x,k_1)W_{\text{on-shell}}^{\alpha_2\beta_2}(x,k_2)K^\mu_{\delta_1}K^\nu_{\zeta_1}   K_{1,\gamma_1\alpha_1}K_{2,\gamma_2\alpha_2}K'_{\delta_2\zeta_2}K_{1,\beta_1\eta_1}K_{2,\beta_2\eta_2}\nonumber\\
&&\!\!\!\!\!\!- \frac12 W_{\text{on-shell}}^{\alpha_1\beta_1}\left(x,k\right)W_{\text{on-shell}}^{\alpha_2\beta_2}(x,k') K'_{\gamma_2\alpha_2}K_{1,\delta_1\zeta_1}K_{2,\delta_2\zeta_2} K'_{\beta_2\eta_2}\left(K^\nu_{\gamma_1}K^\mu_{\alpha_1}K_{\beta_1\eta_1}+K^\mu_{\gamma_1}K^\nu_{\beta_1}K_{\gamma_1\alpha_1}\right)\bigg]\;.\quad\;\;\label{eq:coll_local}
\end{eqnarray}

As a next step, we want to express the collision kernel in extended phase space. In order to do this, we first express the on-shell part of the noninteracting
Wigner function in the collision term via the distribution function in extended phase space as
\begin{equation}
W^{\mu\nu}_{\text{on-shell}}(x,k)=\int \d S(k) h^{\mu\nu}(k,\s)f(x,k,\s)\;,\label{eq:W_through_h}
\end{equation}
where 
\begin{equation}
    h^{\mu\nu}(k,\s)\coloneqq \frac13 K^{\mu\nu}+\frac{i}{2}\epsilon^{\mu\nu\alpha\beta}\frac{k_\alpha}{m}\s_\beta + K^{\mu\nu}_{\alpha\beta} \s^{\alpha} \s^{\beta}\;.
\end{equation}
Note that in this form only the components of the Wigner function that are orthogonal to the momentum have been considered, which is justified since all Wigner functions that appear in Eq. \eqref{eq:coll_local} are contracted with projectors orthogonal to their respective momenta.
Similarly, we define the second-rank tensor 
\begin{equation}
    H^{\mu\nu}(k,\s)\coloneqq \frac13 K^{\mu\nu}+\frac{i}{2}\epsilon^{\mu\nu\alpha\beta}\frac{k_\alpha}{m}\s_\beta +\frac58 K^{\mu\nu}_{\alpha\beta} \s^{\alpha} \s^{\beta}\;,
\end{equation}
such that we can express the collision kernel in extended phase space defined in Eq. \eqref{C_M_def_new} as $\mathfrak{C}\coloneqq H_{\nu\mu}(k,\s)\mathcal{C}^{\mu\nu}$.
Introducing the phase-space measure
\begin{equation}
\d \Gamma_i \coloneqq  \d K_i \d S_i (k_i) \;,\qquad \d K_i\coloneqq \frac{\d^3 \mathbf{k}_i}{(2\pi\hbar)^3k^0_i}
\end{equation}
and using the fact that
\begin{equation}
    \int \d S(k) h^{\mu\nu}(k,\s) = \int \d S(k) H^{\mu\nu}(k,\s) =K^{\mu\nu},
\end{equation}
we can then write the on-shell part of the collision kernel in extended phase space as 
\begin{equation}
\mathfrak{C}^{\text{local}}_{\text{on-shell}} = \frac12 \int \d \Gamma_1 \, \d \Gamma_2 \, \d \Gamma' \,\d \bar{S}(k) (2\pi\hbar)^4 \delta^{(4)}(k+k'-k_1-k_2) \mathcal{W} \left[f(x,k_1,\s_1)f(x,k_2,\s_2)-f(x,k',\s')f(x,k,\bar{\s})\right] \;,\label{eq:coll_local_final}
\end{equation}
where the local transition rate is given by
\begin{equation}
    \mathcal{W}\coloneqq  \frac{(2\pi\hbar)^3}{32} M^{\gamma_1\gamma_2\delta_1\delta_2}M^{\zeta_1\zeta_2\eta_1\eta_2} h_{\gamma_1\eta_1}(k_1,\s_1) h_{\gamma_2\eta_2}(k_2,\s_2) h_{\zeta_2\delta_2}(k',\s') \left\{H(k,\s),h(k,\bar{\s})\right\}_{\zeta_1\delta_1}\;. \label{eq:W}
\end{equation}
In this expression, the curly brackets denote the anticommutator. Note also the similarity of Eq. \eqref{eq:W} to its spin-1/2 counterpart given in Ref. \cite{Wagner:2022amr}. 
Note that a similar result has already been obtained in Ref. \cite{Mrowczynski:1992hq}, where mean-field effects were included as well, but nonlocal contributions to the collision term were omitted.

\subsection{Nonlocal collisions}
In order to obtain the nonlocal collision term, we have to evaluate the $u_1,\,u_2$-derivatives appearing in Eq. \eqref{eq:coll_int_full}, which we achieve by integrating by parts.
The result can be split into three contributions, which we label with Roman numbers below. First, the term with the $u_1,\,u_2$-derivatives acting on the projectors $(K\pm U_1/2\pm U_2/2)^{\mu\nu}$ is evaluated using the relation
\begin{equation}
    \frac{\partial}{\partial u_\lambda} \left(K\pm\frac{U}{2}\right)^{\mu\nu}\bigg|_{u=0}=- \frac{\partial}{\partial u_\lambda} E^{\mu\nu}\left(k\pm\frac{u}{2}\right)\bigg|_{u=0}= \mp\frac{1}{2m^2} g^{\lambda(\mu}k^{\nu)}\pm \frac{k^\lambda k^\mu k^\nu}{m^4}\;,\label{eq:deriv_proj}
\end{equation}
and the result reads
\begin{eqnarray}
   \mathcal{C}^{\text{nonlocal},\,\mu\nu}_{\perp,\text{on-shell,I}}&=& \frac{i\hbar}{2m^2}\frac{(2\pi\hbar)^7}{32} \int \d K_1\, \d K_2\, \d K' M^{\gamma_1\gamma_2 \delta_1\delta_2} M^{\zeta_1\zeta_2 \eta_1\eta_2}\delta^{(4)}(k+k'-k_1-k_2) (K^{\rho\mu}k^\alpha K^{\nu\beta}-K^{\mu\alpha}K^{\rho\nu}k^\beta) \nonumber\\
    &\times& \partial_\rho \bigg[  W^{\alpha_1\beta_1}_{\text{on-shell}}(x,k_1)W^{\alpha_2\beta_2}_{\text{on-shell}}(x,k_2)K_{1,\beta_1\eta_1}K_{1,\gamma_1\alpha_1}K_{2,\beta_2\eta_2}K_{2,\gamma_2\alpha_2}K'_{\delta_2\zeta_2}g_{\alpha\delta_1}g_{\beta\zeta_1}  \nonumber\\
    &-&\frac12 W^{\alpha_2\beta_2}_{\text{on-shell}}(x,k')W^{\alpha_1\beta_1}(x,k)_{\text{on-shell}} K_{1,\delta_1\zeta_1}K_{2,\delta_2\zeta_2}K'_{\gamma_2\alpha_2}K'_{\beta_2\eta_2}\left(g_{\alpha\alpha_1}g_{\beta\gamma_1} K_{\beta_1\eta_1}+g_{\alpha\eta_1}g_{\beta\beta_1}K_{\gamma_1\alpha_1}\right)   \bigg]\;.\label{eq:coll_nonlocal_1_prelim}
\end{eqnarray}
Translating this equation into extended phase space, we find
\begin{eqnarray}
\mathfrak{C}^{\text{nonlocal}}_{\text{on-shell,I}}&=& -\frac{i\hbar}{2m^2}\frac{(2\pi\hbar)^7}{32} \int \d \Gamma_1\, \d \Gamma_2\, \d \Gamma' M^{\gamma_1\gamma_2 \delta_1\delta_2} M^{\zeta_1\zeta_2 \eta_1\eta_2}\delta^{(4)}(k+k'-k_1-k_2)  h_{\gamma_1\eta_1}(k_1,\s_1) h_{\gamma_2\eta_2}(k_2,\s_2)  \nonumber\\
    &\times& h_{\delta_2\zeta_2}(k',\s')\left[H^{\rho}_{\;\;\delta_1}(k,\s)k_{\zeta_1}-k_{\delta_1}H_{\zeta_1}^{\;\;\rho}(k,\s)\right] \partial_\rho \left[f(x,k_1)f(x,k_2)-\frac12 f(x,k')f(x,k)\right]\;.\label{eq:coll_nonlocal_1}
\end{eqnarray}
Here we used the symmetries of $M$ (cf. Appendix \ref{app:scattering}) as well as the fact that the distribution functions are spin-independent at zeroth order in $\hbar$. 

The second contribution is obtained from the $u_1,\, u_2$-derivatives acting on the other projectors that differ between the gain and loss terms, yielding
\begin{eqnarray}
 \mathcal{C}^{\text{nonlocal},\mu\nu}_{\perp,\text{on-shell,II}}&=& \frac{i\hbar}{2m^2}\frac{(2\pi\hbar)^7}{32} \int \d K_1\, \d K_2\, \d K' M^{\gamma_1\gamma_2 \delta_1\delta_2} M^{\zeta_1\zeta_2 \eta_1\eta_2}\delta^{(4)}(k+k'-k_1-k_2)\nonumber\\
    &\times& \bigg\{ \left[\partial_{\rho} W^{\alpha_1\beta_1}_{\text{on-shell}}(x,k_1)\right]W_{\text{on-shell},\gamma_2\eta_2}(x,k_2)\delta^\rho_{\alpha_1}K^{\mu}_{\delta_1}K^\nu_{\zeta_1} K'_{\delta_2\zeta_2}K_{1,\beta_1[\eta_1}k_{1,\gamma_1]}\nonumber\\
    &&+\left[\partial_{\rho} W^{\alpha_2\beta_2}_{\text{on-shell}}(x,k_2)\right]W_{\text{on-shell},\delta_1\zeta_1}(x,k_2)\delta^\rho_{\alpha_2}K^{\mu}_{\delta_1}K^\nu_{\zeta_1}K'_{\delta_2\zeta_2}K_{2,\beta_2[\eta_2}k_{2,\gamma_2]}\nonumber\\
    &-&\frac12 \left[\partial_\rho W^{\alpha_2\beta_2}_{\text{on-shell}}(x,k')\right]W^{\alpha_1\beta_1}_{\text{on-shell}}(x,k) K_{1,\delta_1\zeta_1}K_{2,\delta_2\zeta_2}\left[K^{\mu}_{\alpha_1}K^\nu_{\gamma_1} \left(K_{\beta_1\eta_1}\delta^\rho_{\alpha_2}K'_{\beta_2[\eta_2}k'_{\gamma_2]}\right.\right. \nonumber\\
    &&\left. +K'_{\gamma_2\alpha_2}K'_{\beta_2\eta_2}\delta^\rho_{\beta_1}k_{\eta_1}\right)\left.+K^{\mu}_{\eta_1}K^\nu_{\beta_1} \left(K_{\gamma_1\alpha_1}\delta^\rho_{\alpha_2}K'_{\beta_2[\eta_2}k'_{\gamma_2]}-K'_{\gamma_2\alpha_2}K'_{\beta_2\eta_2}\delta^\rho_{\alpha_1}k_{\gamma_1}\right)\right]\nonumber\\
    &-&\frac12 \left[\partial_\rho W^{\alpha_1\beta_1}_{\text{on-shell}}(x,k)\right]W_{\text{on-shell},\gamma_2\eta_2}(x,k') K_{1,\delta_1\zeta_1}K_{2,\delta_2\zeta_2}\delta^\rho_{\alpha_1} \left(K^\mu_{\eta_1}K^\nu_{\beta_1} k_{\gamma_1}-K^\mu_{\beta_1} K^\nu_{\gamma_1}k_{\eta_1} \right)
    \bigg\}\;.\label{eq:coll_nonlocal_2_prelim}
\end{eqnarray}
Here we used that, since the nonlocal contributions are of order $\mathcal{O}(\hbar)$, we may take the Wigner function to be symmetric, since $W^{\mu\nu}(x,k)=K^{\mu\nu} f^{(0)}(x,k)+\mathcal{O}(\hbar)$.  Contracting Eq. \eqref{eq:coll_nonlocal_1_prelim} with $H_{\nu\mu}(k,\s)$ and using Eq. \eqref{eq:W_through_h} as well as the symmetries of the vertex $M$, we obtain
\begin{eqnarray}
    \mathfrak{C}_{\text{on-shell,II}}^{\text{nonlocal}}&=&  \frac{i\hbar}{2m^2}\frac{(2\pi\hbar)^7}{32}\int \d \Gamma_1\, \d \Gamma_2\, \d \Gamma' \, (2\pi\hbar)^4\delta^{(4)}(k+k'-k_1-k_2)M^{\gamma_1\gamma_2\delta_1\delta_2}M^{\zeta_1\zeta_2\eta_1\eta_2}\nonumber\\
    &\times& \Big\{ f(x,k_2)\left[\partial_\rho f(x,k_1)\right] \left[h_{\;\;\eta_1}^{\rho}(k_1,\s_1)k_{1,\gamma_1}-k_{1,\eta_1}h^{\;\;\rho}_{\gamma_1}(k_1,\s_1)\right] h_{\gamma_2\eta_2}(k_2,\s_2)h'_{\zeta_2\delta_2}(k',\s')H_{\zeta_1\delta_1}(k,\s)\nonumber\\ 
    &&+f(x,k_1)\left[\partial_\rho f(x,k_2)\right] h_{\gamma_1\eta_1}(k_1,\s_1)\left[h_{\;\;\eta_2}^{\rho}(k_2,\s_2)k_{2,\gamma_2}-k_{2,\eta_2}h^{\;\;\rho}_{\gamma_2}(k_2,\s_2)\right] h'_{\zeta_2\delta_2}(k',\s')H_{\zeta_1\delta_1}(k,\s)\nonumber\\ 
    &&-f(x,k)\left[\partial_\rho f(x,k')\right] h_{\gamma_1\eta_1}(k_1,\s_1)h_{\gamma_2\eta_2}(k_2,\s_2)\left[h_{\;\;\delta_2}^{\rho}(k',\s')k'_{\zeta_2}-k'_{\delta_2}h^{\;\;\rho}_{\zeta_2}(k',\s')\right]H_{\zeta_1\delta_1}(k,\s)\nonumber\\ 
    &&-\frac12\left[f(x,k')\partial_\rho f(x,k)-f(x,k)\partial_\rho f(x,k')\right]
    h_{\gamma_1\eta_1}(k_1,\s_1) h_{\gamma_2\eta_2}(k_2,\s_2) \nonumber\\
    &&\quad \times h_{\zeta_2\delta_2}(k',\s')\left[H^{\rho}_{\;\;\delta_1}(k,\s)k_{\zeta_1}-k_{\delta_1}H_{\zeta_1}^{\;\;\rho}(k,\s)\right] 
    \Big\}\;.
\end{eqnarray}
The third contribution consists in the derivatives acting on the Wigner functions and delta functions in the loss term in Eq. \eqref{eq:coll_int_full}. We compute it as
\begin{eqnarray}
    \mathcal{C}^{\text{nonlocal},\,\mu\nu}_{\perp,\text{on-shell,III}} &=& \frac{i\hbar}{4}\frac{(2\pi\hbar)^7}{32} \int \d K_1\, \d K_2 \, \d K' M^{\gamma_1\gamma_2\delta_1\delta_2}M^{\zeta_1\zeta_2\eta_1\eta_2} \delta^{(4)}(k+k'-k_1-k_2) \nonumber\\
    &&\times \left\{\Big[\partial_\rho W_{\text{on-shell},\gamma_2\eta_2}(x,k')\Big]\left[\partial_k^\rho W_{\text{on-shell}}^{\alpha_1\beta_1}(x,k)\right]+\left[\partial_\rho W^{\alpha_1\beta_1}_{\text{on-shell}}(x,k)\right]\Big[\partial_{k'}^\rho W_{\text{on-shell},\gamma_2\eta_2}(x,k')\Big]\right\}\nonumber\\
    &&\times K_{1,\delta_1\zeta_1}K_{2,\delta_2\zeta_2} \left(K^\mu_{\alpha_1} K^\nu_{\gamma_1} K_{\beta_1\eta_1}-K^\mu_{\eta_1} K^\nu_{\beta_1} K_{\gamma_1\alpha_1}\right)\nonumber\\
    &=&\mathcal{O}(\hbar^2)\;.\label{eq:C_nonlocal_3_prelim}
\end{eqnarray}
The second equality follows from the fact that $W_{\text{on-shell}}^{\mu\nu}(x,k)$ is proportional to $K^{\mu\nu}$ at zeroth order, such that this contribution to the nonlocal collision term vanishes up to first order in $\hbar$. This is consistent with the result for spin-1/2 particles found in Ref. \cite{Wagner:2022amr}.

We define the nonlocal shifts
\begin{subequations}\label{eq:def_Delta}
\begin{eqnarray}
    \Delta_1^\mu&\coloneqq& \frac13\frac{1}{\mathcal{W}}\frac{(2\pi\hbar)^3}{32}\frac{i\hbar}{m^2} M^{\gamma_1\gamma_2\delta_1\delta_2}M^{\zeta_1\zeta_2\eta_1\eta_2} \left(h^\mu_{1}{}_{\eta_1}k_{1,\gamma_1}-k_{1,\eta_1}h_{1,\gamma_1}{}^\mu\right)h_{2,\gamma_2\eta_2}h'_{\zeta_2\delta_2}H_{\zeta_1\delta_1} \;,\\
    \Delta_2^\mu&\coloneqq& \frac13 \frac{1}{\mathcal{W}}\frac{(2\pi\hbar)^3}{32}\frac{i\hbar}{m^2} M^{\gamma_1\gamma_2\delta_1\delta_2}M^{\zeta_1\zeta_2\eta_1\eta_2} h_{1,\gamma_1\eta_1}\left(h^\mu_{2}{}_{\eta_2}k_{2,\gamma_2}-k_{2,\eta_2}h_{2,\gamma_2}{}^\mu\right)h'_{\zeta_2\delta_2}H_{\zeta_1\delta_1} \;,\\
    \Delta'^\mu&\coloneqq& \frac13\frac{1}{\mathcal{W}}\frac{(2\pi\hbar)^3}{32}\frac{i\hbar}{m^2} M^{\gamma_1\gamma_2\delta_1\delta_2}M^{\zeta_1\zeta_2\eta_1\eta_2} h_{1,\gamma_1\eta_1}h_{2,\gamma_2\eta_2}\left(h'^\mu{}_{\delta_2}k'_{\zeta_2}-k'_{\delta_2}h'_{\zeta_2}{}^\mu\right)H_{\zeta_1\delta_1} \;,\\
    \Delta^\mu&\coloneqq& \frac13\frac{1}{\mathcal{W}}\frac{(2\pi\hbar)^3}{32}\frac{i\hbar}{m^2} M^{\gamma_1\gamma_2\delta_1\delta_2}M^{\zeta_1\zeta_2\eta_1\eta_2} h_{1,\gamma_1\eta_1}h_{2,\gamma_2\eta_2}h'_{\zeta_2\delta_2}\left(H^\mu{}_{\delta_1}k_{\zeta_1}-k_{\delta_1}H_{\zeta_1}{}^\mu\right) \;,
\end{eqnarray}
\end{subequations}
where we used the shortand notation $h_1\coloneqq h(k_1,\s_1)$ (and analogously for $h_2$, $h'$ and $H$). Note again the similarity of Eqs. \eqref{eq:def_Delta} to the analogous expression for spin-1/2 particles reported in Eqs. (100) of Ref. \cite{Wagner:2022amr}. With these definitions, the nonlocal collision term, defined as the sum of the individual contributions, reads
\begin{eqnarray}
    \mathfrak{C}_{\text{on-shell}}^{\text{nonlocal}} &=&\frac12 \int \d \Gamma_1 \, \d \Gamma_2 \, \d \Gamma' \,\d \bar{S}(k)\, (2\pi\hbar)^4 \delta^{(4)}(k+k'-k_1-k_2) \mathcal{W} \nonumber\\
    &\times&\left[f(x,k_2)\left(\Delta_1^\mu-\Delta^\mu\right)\partial_\mu f(x,k_1)+f(x,k_1)\left(\Delta_2^\mu-\Delta^\mu\right)\partial_\mu f(x,k_2)-f(x,k)\left(\Delta'^\mu-\Delta^\mu\right)\partial_\mu f(x,k')\right]\;, \;\;\label{eq:coll_nonlocal}
\end{eqnarray}
where we introduced a spurious integration over $\bar{\s}$ in order to be able to combine Eqs. \eqref{eq:coll_local_final} and \eqref{eq:coll_nonlocal} in the next subsection.

\subsection{Full collision term}
Adding up the results \eqref{eq:coll_local_final} and \eqref{eq:coll_nonlocal}, we arrive at the main result of this work,
\begin{eqnarray}
k\cdot \partial f(x,k,\s)&=&\frac12 \int \d \Gamma_1 \, \d \Gamma_2 \, \d \Gamma' \,\d \bar{S}(k)\, (2\pi\hbar)^4 \delta^{(4)}(k+k'-k_1-k_2) \mathcal{W}\nonumber\\
&\times&\left[f(x+\Delta_1-\Delta,k_1,\s_1)f(x+\Delta_2-\Delta,k_2,\s_2)-f(x+\Delta'-\Delta,k',\s')f(x,k,\bar{\s})\right] \;,\label{eq:coll_final}
\end{eqnarray}
where $\mathcal{W}$ is defined in Eq. \eqref{eq:W}, while the nonlocal shifts $\Delta_1^\mu$, $\Delta_2^\mu$, $\Delta'^\mu$ and $\Delta^\mu$ have been introduced in Eqs. \eqref{eq:def_Delta}.
Furthermore, we used the fact that the nonlocal shifts are of first order in $\hbar$, such that we may interpret the local and nonlocal contributions as the zeroth and first term in a Taylor series, respectively.

At this point it should be noted that it is in general not possible to formulate a weak equivalence principle for the collision kernel as done in Ref. \cite{Weickgenannt:2021cuo}, i.e., remove the integration over $\bar{\s}$. The reason for this lies in the appearance of terms of second order in the spin variables, and thus the inequivalence of $h^{\mu\nu}(k,\bar{\s})$ and $H^{\mu\nu}(k,\s)$ at this order. 
Nevertheless, in cases where only terms up to first order in the spin vectors are needed, a weak equivalence principle can be formulated, cf. Secs. \ref{sec:eq} and \ref{sec:spin_hydro}.

\section{Equilibrium}
\label{sec:eq}
In addition to conserved scalar quantities such as baryon number or electric charge, the so-called summational invariants in binary elastic collisions are the four-momentum $k^\mu$ and the total angular momentum $J^{\mu\nu}\coloneqq x^{[\mu}k^{\nu]}+\hbar\Sigma_\s^{\mu\nu}$, where the first term constitutes an orbital part and $\Sigma_\s^{\mu\nu}\coloneqq -\epsilon^{\mu\nu\alpha\beta} \frac{k_\alpha}{m} \s_\beta$ denotes the dipole-moment tensor. Since local equilibrium is defined by maximizing the entropy, which in kinetic theory is equivalent to a vanishing collision term, the local-equilibrium distribution function has to consist of the conserved quantities. Furthermore, since consider the low-density approximation, the local-equilibrium distribution function has to be of Maxwell-J{\"u}ttner form, i.e., an exponential. 
Thus, we may write the local-equilibrium distribution function as~\cite{Weickgenannt:2020aaf,Weickgenannt:2021cuo}
\begin{equation}
f_{\text{eq}}(x,k,\s)=\frac{1}{(2\pi\hbar)^3}\exp\left[\alpha(x)-\beta^\mu(x) k_\mu +\frac{\hbar}{2}\Omega_{\mu\nu}(x)\Sigma_\s^{\mu\nu} \right]\;,
\label{Ansatz_f}
\end{equation}
where $\alpha,\,\beta^\mu$ and $\Omega^{\mu\nu}$ are Lagrange multipliers, cf. also Refs.~\cite{Weickgenannt:2020aaf,Weickgenannt:2019dks,Weickgenannt:2022qvh,Weickgenannt:2022zxs}. Note that $\beta^\mu \coloneqq  b^\mu +\Omega^{\mu\nu}x_\nu$, where $b^\mu$ is the Lagrange multiplier for the four-momentum and $\Omega^{\mu\nu}$ the Lagrange multiplier for the total angular momentum \cite{Weickgenannt:2020aaf}. It should be stressed that with this form of the local-equilibrium distribution function there is no tensor polarization in local equilibrium, since the tensor polarization is related to the components of the distribution function that are bilinear in the spin vector. Thus, when taking vector particles as fundamental degrees of freedom, to first order in $\hbar$ the tensor polarization is a purely dissipative effect, cf. Ref. \cite{Wagner:2022gza}.

Expanding the right-hand side of \eq\eqref{eq:coll_final} up to first order in $\hbar$, the collision term reads
\begin{eqnarray}
\mathfrak{C}_{\text{on-shell}}[f_{\text{eq}}]&=&-\int \d \Gamma_1\, \d\Gamma_2 \, \d \Gamma'\, \d \bar{S}(k)\mathcal{W} e^{2\alpha -\beta\cdot(k_1+k_2)}\Big[\partial_\mu \alpha (\Delta^\mu +\Delta'^\mu -\Delta_1^\mu -\Delta_2^\mu)\nonumber\\
&&+\partial_\mu\beta_\nu \left(\Delta_1^\mu k_1^\nu +\Delta_2^\mu k_2^\nu -\Delta^\mu k^\nu -\Delta'^\mu k'^\nu\right) -\frac{\hbar}{2}\Omega_{\mu\nu}\left(\Sigma_{\s_1}^{\mu\nu}+\Sigma_{\s_2}^{\mu\nu}-\Sigma_{\bar{\s}}^{\mu\nu}-\Sigma_{\s'}^{\mu\nu}\right)\Big]  \;.\label{eq:coll_eq_1}
\end{eqnarray}
To proceed, we would like to use the conservation of the total angular momentum, $J^{\mu\nu}_1+J^{\mu\nu}_2-J^{\mu\nu}-J'^{\mu\nu}=0$, which however is not immediately possible because the spin variable after the collision is $\bar{\s}$, and not $\s$. 
As mentioned at the end of Sec. \ref{sec:coll}, in this case it is possible to formulate a weak equivalence principle as done in Ref. \cite{Weickgenannt:2020aaf}. There the goal is to eliminate the integration over $\bar{\s}$ in order to obtain a form of the collision term with standard gain and loss terms. 
Since only spin-integrated quantities are physical, we aim to find a new distribution function $\tilde{f}(x,k,\s)$ as well as a collision term (omitting the subscript ``on-shell'') $\tilde {\mathfrak{C}}[\tilde{f}]$ such that
\begin{subequations}\label{weak_equiv}
\begin{eqnarray}
k\cdot \partial \tilde{f}(x,k,\s)&=& \tilde{\mathfrak{C}}[\tilde{f}]\;,\\
\int \d S(k) b \,\tilde{Q}   &=& \int \d S(k) b\, Q  \;,\label{weak_equiv_int}
\end{eqnarray}
\end{subequations}
where $b\inmath \{1,\s^\mu ,K^{\mu\nu}_{\alpha\beta} \s^\alpha \s^\beta  \}$, $\tilde{Q}\inmath \{ \tilde{f},\tilde{\mathfrak{C}}\}$ and $Q\inmath \{ f,\mathfrak{C}\}$.
Analogous to the argumentation put forward in Ref. \cite{Weickgenannt:2021cuo}, we obtain $\tilde{f}=f$ from Eq. \eqref{weak_equiv_int}. 
For parity-conserving interactions it holds that
\begin{equation}
\int \d S_i (k_i) \mathcal{W}\, \s_i^\mu =0\;,\qquad \int \d S_i (k_i)  \d S_j (k_j) \mathcal{W} \,\s_i^\mu K^{\nu\lambda}_{\alpha\beta} \,\s_j^\alpha \s_j^\beta=0\;,\label{W_odd_0}
\end{equation}
where $\s_i,\s_j\inmath \{\s_1, \s_2, \s', \s,\bar{\s}\}$, since the quantities in Eq. \eqref{W_odd_0} transform as pseudo tensors under parity, while the only tensor structures at our disposal are given by an odd number of powers of momentum, which transform as tensors. Thus, the only term that is nonzero after integration over spin space is the one linear in $\s$. Making use of the equalities
\begin{subequations}\label{ids}
\begin{eqnarray}
\int\d \bar{S}(k)  \left[H^{\mu\alpha}(k,\s) h_{\alpha}^{\;\;\nu}(k,\bar{\s})+h^{\mu\alpha}(k,\bar{\s}) H_{\alpha}^{\;\;\nu}(k,\s)\right] &=&2H^{\mu\nu}(k,\s)\;,\label{id_1}\\
\int \d S (k) \s^\rho \d \bar{S}(k)\bar{\s}^\sigma \left[H^{\mu\alpha}(k,\s) h_{\alpha}^{\;\;\nu}(k,\bar{\s})+h^{\mu\alpha}(k,\bar{\s}) H_{\alpha}^{\;\;\nu}(k,\s)\right] &=&2 \int \d S (k) \s^\rho \s^\sigma H^{\mu\nu}(k,\s)\label{id_2}\;,
\end{eqnarray} 
\end{subequations}
we may replace $\bar{\s}$ with $\s$ and remove the $\d\bar{S}$-integral in Eq. \eqref{div_S} while redefining the transition rate as
\begin{equation}
\widetilde{\mathcal{W}}\coloneqq  \frac{(2\pi\hbar)^3}{16} M^{\gamma_1\gamma_2\delta_1\delta_2}M^{\zeta_1\zeta_2\eta_1\eta_2} H_{\gamma_1\eta_1}(k_1,\s_1) H_{\gamma_2\eta_2}(k_2,\s_2) H_{\delta_2\zeta_2}(k',\s') H_{\delta_1\zeta_1}(k,\s) \;, 
\end{equation}
where we used that $h^{\mu\nu}(k,\s)=H^{\mu\nu}(k,\s)$ to linear order in $\s$ (which are the only relevant terms for this case).
Then, we are able to employ the conservation of the total angular momentum to obtain the modified collision term
\begin{eqnarray}
\tilde{\mathfrak{C}}[f_{\text{eq}}]&=&-\int \d \Gamma_1 \,\d\Gamma_2\, \d \Gamma' \widetilde{\mathcal{W}}\, e^{-\beta\cdot(k_1+k_2)}\Big[\partial_\mu \alpha (\Delta^\mu +\Delta'^\mu -\Delta_1^\mu -\Delta_2^\mu)\nonumber\\
&&+\frac12 \partial_{(\mu}\beta_{\nu)} \left(\Delta_1^\mu k_1^\nu +\Delta_2^\mu k_2^\nu -\Delta^\mu k^\nu -\Delta'^\mu k'^\nu\right) -\frac{\hbar}{2}(\Omega_{\mu\nu}-\varpi_{\mu\nu})\left(\Sigma_{\s_1}^{\mu\nu}+\Sigma_{\s_2}^{\mu\nu}-\Sigma_{\s}^{\mu\nu}-\Sigma_{\s'}^{\mu\nu}\right)\Big]\;,\label{eq:coll_eq_2}
\end{eqnarray}
which is equivalent to the original equilibrium collision term \eqref{eq:coll_eq_1}. Here the thermal vorticity is defined as $\varpi^{\mu\nu}\coloneqq -\frac12 \partial^{[\mu}\beta^{\nu]}$.
Thus, as in the case of spin-1/2 particles \cite{Weickgenannt:2021cuo}, the collision term vanishes to this order if $\partial_\mu \alpha=0$, $\Omega^{\mu\nu}=\varpi^{\mu\nu}$ and $\partial^{(\mu}\beta^{\nu)}=0$,
which are the conditions for global equilibrium as well.

\section{Spin hydrodynamics from kinetic theory}
\label{sec:spin_hydro}

In this section we shortly discuss the implications of the kinetic theory developed previously to the formulation of spin hydrodynamics for massive spin-1 particles. Such a theory was recently derived for spin-1/2 particles in Ref. \cite{Weickgenannt:2022zxs} by using the method of moments to obtain hydrodynamic equations of motion from the Boltzmann equation.
The main difference between spin- and standard hydrodynamics lies in the fact that, in addition to the equations of motion for the energy-momentum tensor and other conserved currents, one has to supply an evolution equation for the spin tensor $S^{\lambda,\mu\nu}$ as well~\cite{Florkowski:2017ruc,Florkowski:2017dyn}. The form of this equation of motion depends on the so-called pseudogauge~\cite{Hehl:1976vr,Speranza:2020ilk}, which has recently been discussed for interacting spin-1/2 and spin-1 particles in Ref. \cite{Weickgenannt:2022jes}. It was found that for the case of spin-1/2 particles the so-called Hilgevoord-Wouthuysen (HW) spin tensor is conserved for free fields or in global equilibrium as well as for a purely local collision term~\cite{Speranza:2020ilk,Weickgenannt:2020aaf}. We will show in the following that this property of the HW spin tensor also holds for spin-1 particles.

We consider the spin tensor in the HW pseudogauge \cite{ Speranza:2020ilk,Weickgenannt:2022jes}
\begin{equation}
S^{\lambda,\mu\nu}_{HW}=\int \d \Gamma k^\lambda \left(\Sigma_\s^{\mu\nu}-\frac{\hbar}{3m^2} k^{[\mu} \partial^{\nu]} \right) f (x,k,\s)+\mathcal{O}(\hbar^2)\;,\label{eq:kin_S_HW}
\end{equation}
where off-shell contributions are neglected due to the low-density approximation~\cite{DeGroot:1980dk}.
Using \eq\eqref{Boltzmann_f} with the local collision term given by \eq\eqref{eq:coll_local_final}, we obtain the following equation of motion for the spin tensor
\begin{eqnarray}
    \partial_\lambda S^{\lambda,\mu\nu}_{HW}\Big|_{\text{local}}&= &\int \d\Gamma\, \Sigma_\s^{\mu\nu} \mathfrak{C}^{\text{local}}_{\text{on-shell}}[f] \nonumber\\
&=& \frac12 \int \d\Gamma \,\d \Gamma_1 \,\d\Gamma_2\, \d \Gamma'\, \d \bar{S}(k)\, \Sigma_\s^{\mu\nu} (2\pi\hbar)^4\delta^{(4)}(k+k'-k_1-k_2)\, \mathcal{W}\nonumber\\
&&\times\left[ f(x,k_1,\s_1)f(x,k_2,\s_2) - f(x,k',\s')f(x,k,\bar{\s})\right]\;.\label{div_S}
\end{eqnarray}
In order to proceed, it is important that the right-hand side of Eq. \eqref{div_S} takes on the standard form of gain and loss terms, i.e., $\bar{\s}$ has to be replaced by $\s$.
Because the dipole-moment tensor $\Sigma_\s^{\mu\nu}$ is linear in $\s$, it is possible to employ the weak equivalence principle as shown in Sec. \ref{sec:eq}, i.e., replace $\mathcal{W}$ by $\widetilde{\mathcal{W}}$ and remove the $\d \bar{S}$-integration. 
Since $\widetilde{\mathcal{W}}$ is manifestly symmetric under the exchanges $(k_1,\s_1)\leftrightarrow (k_2,\s_2)$, $(k,\s)\leftrightarrow (k',\s')$ and $[(k_1,\s_1);(k_2,\s_2)]\leftrightarrow [(k,\s);(k',\s')]$, the summational conservation of spin in local collisions, $\Sigma_\s^{\mu\nu}+\Sigma_{\s^\prime}^{\mu\nu}=\Sigma_{\s_1}^{\mu\nu}+\Sigma_{\s_2}^{\mu\nu}$, implies that in this case the HW spin tensor is a conserved quantity. On the other hand, for nonlocal collisions the dipole-moment tensor would not be a collisional invariant and, therefore, the HW spin tensor is in general not conserved.

\section{Conclusions}
\label{sec:conclusion}
In this paper, we have derived the collision term for massive spin-1 particles from the Wigner-function formalism up to first order in $\hbar$, following the method outlined in Ref. \cite{DeGroot:1980dk} and recently employed for spin-1/2 particles in Refs. \cite{Weickgenannt:2021cuo,Weickgenannt:2020aaf}. 
Both local and nonlocal contributions have been computed in a covariant fashion, resulting in similar expressions to those derived in Ref. \cite{Wagner:2022amr} for spin-1/2 particles.

We find that it is in general not possible to formulate a weak equivalence principle for the collision term as done in Ref. \cite{Weickgenannt:2021cuo} in order to obtain the standard form of the gain and loss terms. However, when the distribution function is at most of first order in the phase-space spin variable, such a replacement is possible, allowing one to establish the usual form of the local-equilibrium distribution function and to show that the dipole-moment tensor $\Sigma_\s^{\mu\nu}$ is conserved in local collisions. 
Furthermore, we do not find contributions of second order in the spin variable in local equilibrium, suggesting that all effects related to the spin alignment of vector mesons are either of dissipative origin \cite{Wagner:2022gza} or of higher order in $\hbar$.

The theory developed in this paper can be employed to evaluate the spin alignment of particles such as the $\phi$-mesons, which can then be compared to experimental results. As a first step in this direction, Ref. \cite{Wagner:2022gza} evaluated the hydrodynamic Navier-Stokes limit of the kinetic theory presented here to derive a  relation between the spin alignment of vector mesons and the shear stress of the medium.
In order to refine these results, one can derive second-order dissipative spin-1 hydrodynamics by using the method of moments, as it has recently been carried out for the spin-1/2 case \cite{Weickgenannt:2022zxs,Weickgenannt:2022qvh}.

\appendix

\section*{Acknowledgments}

The work of D.W. and N.W.\ is supported by the
Deutsche Forschungsgemeinschaft (DFG, German Research Foundation)
through the Collaborative Research Center CRC-TR 211 ``Strong-interaction matter
under extreme conditions'' -- project number 315477589 - TRR 211 and by the State of Hesse within the Research
Cluster ELEMENTS (Project ID 500/10.006). D.W.\ acknowledges support by the Studienstiftung des deutschen Volkes 
(German Academic Scholarship Foundation) as well as the support through a grant of the
Ministry of Research, Innovation and Digitization, CNCS
- UEFISCDI, project number PN-III-P1-1.1-TE-2021-
1707, within PNCDI III.
N.W.\ acknowledges support by the German National Academy of Sciences Leopoldina through the Leopoldina fellowship program with funding code  LPDS 2022-11. This research was supported in part by the National Science Foundation under Grant No. NSF PHY-1748958.

\appendix 

\section{Rewriting expectation values}
\label{app:rewriting}
In this appendix we show how to express the expectation value of an arbitrary operator in terms of the ``in''-Wigner function \eqref{W_in_def}.
Consider a general operator $\hat{O}$ acting in the Hilbert space. Inserting Eq. \eqref{completeness}, we can express its statistical average as
\begin{equation}
\avg{\hat{O}}\coloneqq \mathrm{Tr}\,\hat{\rho}\,\hat{O} =\sum_{n=0}^\infty \frac{1}{(n!)^2} \sum_{\lambda^n,\lambda'^n}\momint{k}{n}{}\momint{k'}{n}{} \prescript{}{\in}{\bra{k^n;\lambda^n}} \hat{O}\ket{k'^{n};\lambda'^n}_\in \prescript{}{\in}{\bra{k'^n;\lambda'^n}} \hat{\rho} \ket{k^n;\lambda^n}_\in \;.\label{O_exp}
\end{equation}
Note that, to arrive at this expression, we assume that the density matrix commutes with the number operator of ``in''-state particles \cite{DeGroot:1980dk}.
The next task is to express the matrix element of the density matrix through expectation values of bilinear products of creation and annihilation operators. For this we first compute, using the cyclicity of the trace,
\begin{eqnarray}
\avg{\hat{a}^{\dagger}_\in (k^n,\lambda^n) \hat{a}_\in (k'^n,\lambda'^n)}&=&\mathrm{Tr}\, \hat{a}_\in (k'^n,\lambda'^n)\hat{\rho}\, \hat{a}^\dagger_\in (k^n,\lambda^n) \nonumber\\
&=&\sum_{k=0}^\infty \frac{1}{k!}\sum_{\sigma^n} \momint{p}{k}{} \prescript{}{\in}{\bra{p^k,k'^n;\sigma^k,\lambda'^n}}  \hat{\rho}  \ket{p^k,k^n;\sigma^k,\lambda^n}_\in\;.
\end{eqnarray}
The inversion of this relation, proven in Ref. \cite{DeGroot:1980dk}, gives
\begin{eqnarray}
\prescript{}{\in}{\bra{k'^n;\lambda'^n}}\hat{\rho}\ket{k^n;\lambda^n}_\in &=& \sum_{m=0}^\infty \frac{(-1)^m}{(m!)^2}\sum_{\sigma^m,\sigma'^m}\momint{p}{m}{}\momint{p'}{m}{} \prescript{}{\in}{\braket{p^m;\sigma^m}{p'^m;\sigma'^m}_\in}\nonumber\\
&&\times \avg{ \hat{a}^\dagger_\in(k^{n},p^m,\lambda^n,\sigma^m)\hat{a}_\in(k'^{n},p'^m,\lambda'^n,\sigma'^m) }\;,
\end{eqnarray}
which we may insert into Eq. \eqref{O_exp} to obtain
\begin{eqnarray}
\avg{\hat{O}}&=&\sum_{n=0}^\infty \frac{1}{(n!)^2}\sum_{\lambda^n,\lambda'^n} \momint{k}{n}{}\momint{k'}{n}{} \prescript{}{\in}{\bra{k^n;\lambda^n}} \hat{O}\ket{k'^{n};\lambda'^n}_\in\nonumber\\
&&\times\sum_{m=0}^\infty \frac{(-1)^m}{(m!)^2}\sum_{\sigma^m,\sigma'^m}\momint{p}{m}{}\momint{p'}{m}{} \prescript{}{\in}{\braket{p^m;\sigma^m}{p'^m;\sigma'^m}_\in}\nonumber\\
&&\times \avg{ \hat{a}^\dagger_\in(k^{n},p^m,\lambda^n,\sigma^m)\hat{a}_\in(k'^{n},p'^m,\lambda'^n,\sigma'^m) }\;.
\end{eqnarray}
Introducing a new summation index, $j\coloneqq n+m$, and using the fact that $\sum_{n=0}^\infty \sum_{m=0}^\infty \equiv \sum_{j=0}^\infty \sum_{m=0}^j$, we arrive at
\begin{equation}
\avg{\hat{O}}=\sum_{j=0}^\infty \left(\frac{1}{j!}\right)^2 \sum_{\lambda^j,\lambda'^j} \momint{k}{j}{} \momint{k'}{j}{} \prescript{}{\in}{\left\llangle k^j;\lambda^j \right|\hat{O}\left| k'^j;\lambda'^j \right\rrangle_\in} \avg{\hat{a}^\dagger_\in(k^j,\lambda^j)\hat{a}_\in(k'^j,\lambda'^j)}\;,\label{avg_O_2}
\end{equation}
where we defined the expression (taken to be symmetric under the exchange of primed and unprimed variables)
\begin{equation}
\prescript{}{\in}{\left\llangle k^j;\lambda^j \right|\hat{O}\left| k'^j;\lambda'^j \right\rrangle_\in} \coloneqq  \sum_{m=0}^j (-1)^m \begin{pmatrix} j\\m\end{pmatrix}^2 \prescript{}{\in}{\braket{k^m;\lambda^m}{k'^m;\lambda'^m}_\in} \prescript{}{\in}{\bra{k^{j-m};\lambda^{j-m}}}\hat{O}\ket{k'^{j-m};\lambda'^{j-m}}_\in \;.\label{double_bracket}
\end{equation}
Next we put in the essential assumption of molecular chaos, implying that the expectation value of creation and annihilation operators factorizes pairwise as
\begin{equation}
\avg{\hat{a}^\dagger_\in(k^n,\lambda^n)\hat{a}_\in (k'^m,\lambda'^m)}=\delta_{nm} \sum_{\mathcal{P}}\prod_{j=1}^n \avg{\hat{a}^\dagger_\in(k_j,\lambda_j)\hat{a}_\in (k'_j,\lambda'_j)}\;.\label{mol_chaos_a}
\end{equation}
Here the symbol $\mathcal{P}$ stands for the symmetrization with respect to the primed and unprimed variables, which is necessary because of the bosonic nature of the particles.
In terms of the fields, Eq. \eqref{mol_chaos_a} becomes
\begin{equation}
\avg{\hat{V}^{\dagger\mu_1}_\in(x_1)\cdots \hat{V}^{\dagger\mu_n}_\in(x_n)\hat{V}^{\nu_1}_\in (x'_1)\cdots \hat{V}^{\nu_m}_\in (x'_m)}=\delta_{nm} \sum_{\mathcal{P}}\prod_{j=1}^n \avg{\hat{V}^{\dagger\mu_j}_\in(x_j)\hat{V}^{\nu_j}_\in (x'_j)}\;.\label{mol_chaos_V}
\end{equation}
Inverting the definition of the ``in''-fields \eqref{V_in}, we have
\begin{equation}
\frac{1}{2\pi\hbar^{3/2}} \int \d^4 x e^{\frac{i}{\hbar}k\cdot x}  \epsilon^{*(\lambda)}_\mu (k) \hat{V}^\mu_\in (x)= -\Theta(k^0) \delta(k^2-m^2)  \hat{a}_\in (k,\lambda)\;.
\end{equation}
Inserting this expression into Eq. \eqref{avg_O_2}, we obtain
\begin{equation}
\avg{\hat{O}}=\sum_{n=0}^\infty \frac{1}{n!} \int \d^4 x^n \int \d^4 x'^n  \tilde{O}_{n,\mu_1\nu_1\cdots \mu_n\nu_n}(x^n;x'^n) \prod_{j=1}^n \avg{\hat{V}^{\dagger\mu_j}_{\in}(x_j)\hat{V}^{\nu_j}_{\in}(x'_j)}\;,\label{avg_O_3}
\end{equation}
where
\begin{eqnarray}
\tilde{O}_{n,\mu_1\nu_1\cdots \mu_n \nu_n}(x^n;x'^n)&\coloneqq &\hbar^{-n}\int \frac{\d^4 k^n}{(2\pi\hbar)^{4n}}\int \frac{\d^4 k'^n}{(2\pi\hbar)^{4n}} \sum_{\lambda^n,\lambda'^n} \left[\prod_{j=1}^n e^{-\frac{i}{\hbar} \left(k_j x_j-k'_j x'_j   \right)  } \epsilon^{(\lambda_j)}_{\mu_j} (k_j) \epsilon^{*(\lambda'_j)}_{\nu_j} (k'_j) \right]\nonumber\\
&&\times\prescript{}{\in}{\left\llangle  k^n; \lambda^n\right| \hat{O} \left| k'^n;\lambda'^n \right\rrangle_\in} \;.
\end{eqnarray}

Using the definition of the ``in''-Wigner function \eqref{W_in_def}, we obtain
\begin{equation}
\avg{\hat{V}^{\dagger\mu}_\in \left(x+\frac{v}{2}\right) \hat{V}^\nu_\in \left(x-\frac{v}{2}\right)   }=-\frac{\hbar}{2}\int \d^4 k e^{\frac{i}{\hbar} k\cdot v} W^{\mu\nu}_\in (x,k)\;.\label{W_inv}
\end{equation}
Defining the center and difference variables $\bar{x}_j\coloneqq (x_j+x'_j)/2$ and $v_j\coloneqq x_j-x'_j$, Eq. \eqref{W_inv} in conjunction with Eq. \eqref{avg_O_3} yields
\begin{equation}
\avg{\hat{O}}=\sum_{n=0}^\infty \frac{1}{n!} \int \d^4 \bar{x}^n \int \d^4 \bar{k}^n O_{n,\mu_1\nu_1\cdots \mu_n\nu_n}(\bar{x}^n;\bar{k}^n) \prod_{j=1}^n W_\in^{\mu_j\nu_j} (\bar{x}_j,\bar{k}_j)\;,\label{avg_O_4}
\end{equation}
where we defined
\begin{eqnarray}
O_{n,\mu_1\nu_1\cdots \mu_n \nu_n}(\bar{x}^n;\bar{k}^n)&\coloneqq &\frac{(-1)^n}{2^n(2\pi\hbar)^{4n}}\int \d^4 u^n \sum_{\lambda^n,\lambda'^n} \left[\prod_{j=1}^n e^{\frac{i}{\hbar} u_j\cdot \bar{x}_j  } \epsilon^{(\lambda_j)}_{\mu_j} \left(\bar{k}_j-\frac{u_j}{2}\right) \epsilon^{*(\lambda'_j)}_{\nu_j} \left(\bar{k}_j+\frac{u_j}{2}\right) \right]\nonumber\\
&&\times \prescript{}{\in}{\left\llangle  \bar{k}^n-\frac{u^n}{2}; \lambda^n\right| \hat{O} \left| \bar{k}^n+\frac{u^n}{2};\lambda'^n \right\rrangle_\in} \;.
\end{eqnarray}
Note that in this calculation $\bar{k}$ is the integration variable appearing in Eq. \eqref{W_inv}, we used the emerging delta function $\delta^{(4)}\left(\frac{k_j+k'_j}{2}-\bar{k}_j\right)$ and defined $k_j-k'_j\eqqcolon u_j$.

Equation \eqref{avg_O_4} is the sought-after relation that allows to express the expectation value of any operator in terms of $W_\in^{\mu\nu}$.

\section{Expansion in terms of \texorpdfstring{$W_\in^{\mu\nu}$}{W\_in}}
\label{app:scattering}

In this appendix we show how to use Eq. \eqref{avg_O_4} in order to express both the Wigner function and the collision kernel in terms of $W_\in^{\mu\nu}$.
\subsection{Expansion of \texorpdfstring{$W^{\mu\nu}$}{W} in terms of \texorpdfstring{$W_\in^{\mu\nu}$}{W\_in}}
The (positive-frequency) Wigner function may be expressed as the following ensemble average, 
\begin{equation}
W^{\mu\nu}(x,k)=\avg{ e^{\frac{i}{\hbar}\hat{P}\cdot x} \hat{\Psi}^{\mu\nu} (k) e^{-\frac{i}{\hbar}\hat{P}\cdot x}   }\;,\label{mom_op_W}
\end{equation}
where 
\begin{equation}
\hat{\Psi}^{\mu\nu}(k)\coloneqq -\frac{2}{(2\pi\hbar)^4\hbar} \int \d^4 v e^{-\frac{i}{\hbar}k\cdot v} :\hat{V}^{\dagger\mu}\left(\frac{v}{2}\right) \hat{V}^{\nu} \left(-\frac{v}{2}\right):
\end{equation}
and $\hat{P}$ is the usual momentum operator. Since the ``in''-states are eigenstates of the total momentum, we may replace $e^{-\frac{i}{\hbar}\hat{P}\cdot x}\ket{k^n;\lambda^n}_\in=\prod_{j=1}^n e^{-\frac{i}{\hbar}k_j\cdot x}\ket{k^n;\lambda^n}_\in$. Rethinking the steps that led to Eq. \eqref{avg_O_4}, we observe that the definition of $\bar{x}_j$ will contain an additional term of $x$. Thus we have
\begin{equation}
W^{\mu\nu}(x,k)=\sum_{n=0}^\infty \frac{1}{n!} \int \d^4 \bar{x}^n \int \d^4 \bar{k}^n \Psi^{\mu\nu}_{n,\alpha_1\beta_1\cdots \alpha_n \beta_n}(\bar{x}^n;\bar{k}^n|k) \prod_{j=1}^n W_\in^{\alpha_j\beta_j} (x+\bar{x}_j,\bar{k}_j)\;,\label{avg_W_1}
\end{equation}
where
\begin{eqnarray}
\Psi^{\mu\nu}_{n,\alpha_1\beta_1\cdots \alpha_n\beta_n}(\bar{x}^n;\bar{k}^n|k)&\coloneqq &\frac{(-1)^n}{2^n(2\pi\hbar)^{4n}}\int \d^4 u^n \sum_{\lambda^n,\lambda'^n} \left[\prod_{j=1}^n e^{\frac{i}{\hbar} u_j\cdot \bar{x}_j  } \epsilon^{(\lambda_j)}_{\alpha_j} \left(\bar{k}_j-\frac{u_j}{2}\right) \epsilon^{*(\lambda'_j)}_{\beta_j} \left(\bar{k}_j+\frac{u_j}{2}\right) \right]\nonumber\\
&&\times \prescript{}{\in}{\left\llangle  \bar{k}^n-\frac{u^n}{2}; \lambda^n\right| \hat{\Psi}^{\mu\nu}(k) \left| \bar{k}^n+\frac{u^n}{2};\lambda'^n \right\rrangle_\in} \;.
\end{eqnarray}
Following Ref. \cite{DeGroot:1980dk}, we compute $\Psi^{\mu\nu}_{n,\alpha\beta}$ for $n=0,1$. 

In the case $n=0$, inserting Eq. \eqref{double_bracket} and making use of the completeness of the ``out''-states, we obtain
\begin{eqnarray}
\Psi^{\mu\nu}_{0}(0|k)
&=&- \frac{2}{\hbar}\Theta(k^0) \Theta(k^2)\sum_{m=0}^\infty \frac{1}{m !} \sum_{\lambda^m} \momint{k}{m}{}\bra{ 0} \hat{V}^{\dagger\mu}\left(0\right) \ket{k^m;\lambda^m}_\out \nonumber\\
&&\times\prescript{}{\out}{\bra{k^m;\lambda^m}} \hat{V}^\nu \left(0\right) \ket{0} \delta^{(4)}\left( k +\sum_{\ell=1}^m  k_\ell\right)\;,
\end{eqnarray}
where we employed the fact that the ``out''-states are eigenstates of the momentum as well.
Since due to the Heaviside functions (which are also implied in the momentum integrals) the zero-component of the momenta is always positive, the delta function is always zero, such that
\begin{equation}
\Psi^{\mu\nu}_{0}(0|k)=0\;.
\end{equation}
Similarly, the $n=1$ case yields
\begin{eqnarray}
\Psi_{1,\alpha\beta}^{\mu\nu}(x;p|k)&=&-\frac{1}{2(2\pi\hbar)^4} \int \d^4u \sum_{\lambda,\lambda'} e^{\frac{i}{\hbar} u\cdot x} \epsilon_\alpha^{(\lambda)}\left(p-\frac{u}{2}\right)\epsilon_\beta^{*(\lambda')}\left(p+\frac{u}{2}\right)\nonumber\\
&&\times\left[\prescript{}{\in}{\bra{p-\frac{u}{2};\lambda}}\hat{\Psi}^{\mu\nu} \ket{p+\frac{u}{2};\lambda'}_\in -\prescript{}{\in}{\braket{p-\frac{u}{2};\lambda}{p+\frac{u}{2};\lambda'}}{}_\in \bigg\langle0\bigg|\hat{\Psi}^{\mu\nu} \bigg|0\bigg\rangle\right]\nonumber\;.
\end{eqnarray}
Note that the second term vanishes for the same reasons as $\Psi^{\mu\nu}_{0}$. Inserting a complete set of ``out''-states again, we have
\begin{eqnarray}
\Psi_{1,\alpha\beta}^{\mu\nu}(x;p|k)
&=&\frac{1}{(2\pi\hbar)^4\hbar }   \int \d^4u \sum_{\lambda,\lambda'} e^{\frac{i}{\hbar} u\cdot x} \epsilon_\alpha^{(\lambda)}\left(p-\frac{u}{2}\right)\epsilon_\beta^{*(\lambda')}\left(p+\frac{u}{2}\right)\sum_{m=0}^\infty \frac{1}{m!} \sum_{\sigma^m}\nonumber\\
&&\times  \momint{p'}{m}{} \prescript{}{\in}{\bra{p-\frac{u}{2};\lambda}} \hat{V}^{\dagger\mu}(0) \bigg|p'^m;\sigma^m\bigg\rangle_{\out} \nonumber\\
&&\times\prescript{}{\out}{\bigg\langle p'^m;\sigma^m\bigg|} \hat{V}^\nu(0)  \ket{p+\frac{u}{2};\lambda'}_\in \delta^{(4)}\left(k+\sum_{\ell=1}^m p'_\ell -p\right)\label{Psi_1_1}\;.
\end{eqnarray}
Using the fact that the one-particle ``in''- or ``out''-states are orthogonal, we find that
\begin{equation}
\bigg\langle 0 \bigg| \hat{V}^\nu(0)  \ket{p+\frac{u}{2};\lambda'}_\in = \sqrt{\hbar}\epsilon^{(\lambda')\nu}\left(p+\frac{u}{2}\right) \;,
\end{equation}
which may be used to obtain an explicit expression for the $m=0$-term in Eq. \eqref{Psi_1_1}.
The other terms in the respective sum require that
\begin{equation} p^2=(k+p')^2=2m^2+2k\cdot p'>2m^2\;,  
\end{equation}
which implies including the possibility of creating particles with masses larger than twice the mass of the original vector bosons. This possibility we will neglect, such that only the $m=0$ term in Eq. \eqref{Psi_1_1} contributes, yielding
\begin{eqnarray}
\Psi_{1,\alpha\beta}^{\mu\nu}(x;p|k)&=&\frac{1}{(2\pi\hbar)^4}  \int \d^4u \sum_{\lambda,\lambda'} e^{\frac{i}{\hbar} u\cdot x} \epsilon_\alpha^{(\lambda)}\left(p-\frac{u}{2}\right)\epsilon_\beta^{*(\lambda')}\left(p+\frac{u}{2}\right)\nonumber\\
&&\times \epsilon^{*(\lambda)\mu}\left(p-\frac{u}{2}\right) \epsilon^{(\lambda')\nu}\left(p+\frac{u}{2}\right) \delta^{(4)}\left(k -p\right)\;.
\end{eqnarray}
Truncating the sum in Eq. \eqref{avg_W_1} after the first term (higher orders would lead to nonlinear dependencies of $W^{\mu\nu}$ on $W^{\mu\nu}_\in$) and expanding the ``in''-Wigner function around $x$, we obtain
\begin{eqnarray}
W^{\mu\nu}(x,k)
&=& \int \d^4u \sum_{\lambda,\lambda'}   \epsilon^{*(\lambda)\mu}\left(k-\frac{u}{2}\right)\epsilon_\alpha^{(\lambda)}\left(k-\frac{u}{2}\right) \epsilon_\beta^{*(\lambda')}\left(k+\frac{u}{2}\right)\epsilon^{(\lambda')\nu}\left(k+\frac{u}{2}\right)\nonumber\\
&&\times\left\{ W_\in^{\alpha\beta} (x,k) \delta^{(4)}(u)-i\hbar \left[\partial^\rho_u \delta^{(4)}(u)\right] \partial_\rho W_\in^{\alpha\beta} (x,k) \right\}\nonumber\\
&=& K^\mu_\alpha K^\nu_\beta W_\in^{\alpha\beta} (x,k) -i\hbar \int \d^4u \left(K-\frac{U}{2}\right)^\mu_\alpha \left(K+\frac{U}{2}\right)^\nu_\beta  \left[\partial^\rho_u \delta^{(4)}(u)\right] \partial_\rho W_\in^{\alpha\beta} (x,k)\;, \label{W_1_2}
\end{eqnarray}
where we defined 
\begin{equation}
\left(K\pm \frac{U}{2}\right)^{\mu\nu}\coloneqq g^{\mu\nu}-(k\pm u)^{-2}  \left(k\pm \frac{u}{2}\right)^\mu \left(k\pm \frac{u}{2}\right)^\nu \;.
\end{equation}
Integrating by parts and using
\begin{equation}
\partial_u^\rho \left(K\pm \frac{U}{2}\right)^{\mu\nu}\bigg|_{u=0}=\mp k^{-2}\left( \frac{1}{2} g^{\rho(\mu} k^{\nu)}-\frac{k^\rho k^\mu k^\nu}{k^2}   \right)
\end{equation}
as well as the fact that $k\cdot \partial W_\in^{\mu\nu}(x,k)=0$, we can evaluate the second term in Eq. \eqref{W_1_2} to get
\begin{eqnarray}
W^{\mu\nu}(x,k)&=& K^\mu_\alpha K^\nu_\beta W_\in^{\alpha\beta} (x,k) +i\hbar \partial_u\left[ \left(K-\frac{U}{2}\right)^\mu_\alpha \left(K+\frac{U}{2}\right)^\nu_\beta\right] \partial_\rho W_\in^{\alpha\beta} (x,k)\nonumber\\
&=& K^\mu_\alpha K^\nu_\beta W_\in^{\alpha\beta} (x,k) +\frac{i\hbar}{2}k^{-2} \left(   \partial^{(\mu} k_{\alpha)} K^\nu_\beta - \partial^{(\nu} k_{\beta)} K^\mu_\alpha \right)  W_\in^{\alpha\beta} (x,k)\nonumber\;.
\end{eqnarray}
Remembering that the ``in''-Wigner function corresponds to free particles, we know from the results of free Proca theory \cite{Weickgenannt:2022jes} that $k_\mu W_{S,\in}^{\mu\nu}\sim\mathcal{O}(\hbar^2)$ and $k_\mu W_{A,\in}^{\mu\nu}=(i\hbar/2)\partial^\nu f_K^{(0)}+\mathcal{O}(\hbar^2)$. Thus, we may rewrite this expression up to first order in $\hbar$ as
\begin{equation}
W^{\mu\nu}(x,k)=  K^\mu_\alpha K^\nu_\beta W_\in^{\alpha\beta} (x,k)+\frac{i\hbar}{2}\frac{k^{[\mu}}{k^2}\partial^{\nu]}f_{K,\in}^{(0)}(x,k)   \equiv  W_{\in}^{\mu\nu}(x,k)\label{W_to_Win}\;.
\end{equation}
Manifestly, we may identify the Wigner function with its ``in''-counterpart to this order.
Note that this does not mean that $k\cdot \partial W^{\mu\nu}\sim \mathcal{O}(\hbar^2)$, as the low-density approximation \eqref{W_to_Win} will only be used inside the collision integral.

\subsection{Expansion of \texorpdfstring{$\mathcal{C}^{\mu\nu}$}{C} in terms of \texorpdfstring{$W_\in^{\mu\nu}$}{W\_in}}

Similar to the Wigner function itself, we may also express the expectation value given in Eq. \eqref{def_C}, which determines the right-hand side of the Boltzmann equation, via the ``in''-Wigner functions:
\begin{equation}
\mathcal{C}^{\mu\nu}(x,k)=\sum_{n=2}^\infty \frac{1}{n!} \int \d^4 \bar{x}^n \int \d^4 \bar{k}^n \Phi^{\mu\nu}_{n,\alpha_1\beta_1\cdots \alpha_n\beta_n} (\bar{x}^n;\bar{k}^n|k) \prod_{j=1}^n W_\in^{\alpha_j\beta_j} (x+\bar{x}_j, \bar{k}_j)\;,\label{kin_coll_exp}
\end{equation}
where
\begin{eqnarray}
\Phi^{\mu\nu}_{n,\alpha_1\beta_1\cdots \alpha_n\beta_n}(\bar{x}^n;\bar{k}^n|k)&\coloneqq &\frac{(-1)^n}{2^n(2\pi\hbar)^{4n}}\int \d^4 u^n \sum_{\lambda^n,\lambda'^n} \left[\prod_{j=1}^n e^{\frac{i}{\hbar} u_j\cdot \bar{x}_j  } \epsilon^{(\lambda_j)}_{\alpha_j} \left(\bar{k}_j-\frac{u_j}{2}\right) \epsilon^{*(\lambda'_j)}_{\beta_j} \left(\bar{k}_j+\frac{u_j}{2}\right) \right]\nonumber\\
&&\times \prescript{}{\in}{\left\llangle  \bar{k}^n-\frac{u^n}{2}; \lambda^n\right| \hat{\Phi}^{\mu\nu}(k) \left| \bar{k}^n+\frac{u^n}{2};\lambda'^n \right\rrangle_\in} \;,\label{def_Phi}
\end{eqnarray}
and 
\begin{equation}
\hat{\Phi}^{\mu\nu}(k)\coloneqq -\frac{i}{(2\pi\hbar)^4} \int\d^4 v e^{-\frac{i}{\hbar}k\cdot v} :\left[\hat{V}^{\dagger\mu}\left( \frac{v}{2} \right) \hat{\rho}^\nu \left(-\frac{v}{2}\right)  -\hat{\rho}^{\dagger\mu}\left( \frac{v}{2} \right) \hat{V}^\nu \left(-\frac{v}{2}\right)   \right]:\;.
\end{equation}
Note that the sum in Eq. \eqref{kin_coll_exp} starts at $n=2$, which, following Ref. \cite{DeGroot:1980dk}, can be understood as follows: The $n=0$ term vanishes via the same argumentation as before, since the line of reasoning did not depend on the objects appearing in the two-point function, but rather on their arguments (i.e., $\pm v/2$). The $n=1$ term has to vanish as well since the ``in''-Wigner function represents the distribution of particles without interactions, such that $k\cdot \partial W_\in^{\mu\nu}=0$. 

The kernel \eqref{def_Phi} will be analyzed in the following for $n=2$, corresponding to $2\to 2$ collisions. It should be noted that, considering Eq. \eqref{Boltzmann_f}, only the parts of $\Phi^{\mu\nu}$ orthogonal to $k^\mu$ will contribute to the kinetic equation.

Inserting a complete set of ``out''-states and using the fact that the ``in''-and ``out''-states are momentum eigenstates, we obtain
\begin{eqnarray}
&&\prescript{}{\in}{\bra{k^2-\frac{u^2}{2};\lambda^2}}\hat{\Phi}^{\mu\nu}(k)\ket{k^2+\frac{u^2}{2};\lambda'^2}_\in \nonumber\\
&=&-i \sum_{m=0}^\infty \frac{1}{m!} \sum_{\sigma'^m}\momint{k'}{m}{}\prescript{}{\in}{\bra{k^2-\frac{u^2}{2};\lambda^2}}\bigg[\hat{V}^{\dagger\mu}(0)\bigg|k'^m;\sigma'^m\bigg\rangle_\out  \prescript{}{\out}{\bigg\langle k'^m;\sigma'^m\bigg|} : \hat{\rho}^\nu (0) :\nonumber\\
&&-:\hat{\rho}^{\dagger\mu}(0):\bigg| k'^m;\sigma'^m\bigg\rangle_\out \prescript{}{\out}{\bigg\langle k'^m;\sigma'^m\bigg|} \hat{V}^\nu (0) \bigg]\ket{k^2+\frac{u^2}{2};\lambda'^2}_\in  \delta^{(4)}\left( k  +\sum_{j=0}^m k'_j -k_1-k_2  \right)\nonumber\\
&=&-i \sum_{\sigma'}\momint{k'}{}{}\prescript{}{\in}{\bra{k^2-\frac{u^2}{2};\lambda^2}}\bigg[\hat{V}^{\dagger\mu}(0)\bigg| k';\sigma'\bigg\rangle_\out  \prescript{}{\out}{\bigg\langle k';\sigma'\bigg|} :\hat{\rho}^\nu (0): \nonumber\\
&&-:\hat{\rho}^{\dagger\mu}(0):\bket{k';\sigma'}_\out \prescript{}{\out}{\bbra{k';\sigma'}} \hat{V}^\nu (0) \bigg]\ket{k^2+\frac{u^2}{2};\lambda'^2}_\in  \delta^{(4)}\left( k  +k' -k_1-k_2  \right)\;.\label{Phi_1}
\end{eqnarray}
Note that in the last equality we assumed some conserved charge to be present (e.g. baryon number or electric charge), and assumed only one species of particles. Under these conditions, the only permissible scattering with two outgoing particles is $2\to 2$ scattering.

Our next task consists in evaluating the matrix elements involving the field operators, for which we use the Yang-Feldman equation relating the fields to the ``in''-fields and the source operators $\hat{\rho}^\mu$ \cite{DeGroot:1980dk},
\begin{equation}
\hat{V}^{ \mu}(0)=\hat{V}^{ \mu}_\in (0) +\int \d^4 x\, \Delta_R^{\mu\nu} (-x) \hat{\rho}_\nu (x)\label{Yang_Feldman}\;.
\end{equation}
Here $\Delta_R^{\mu\nu}$ is the (symmetric) retarded vector boson propagator.
Making use of the Fourier decomposition of the ``in''-field operators,
\begin{equation}
\hat{V}^{\mu}_\in (0)=\sqrt{\hbar}\sum_{\sigma'} \momint{k'}{}{} \hat{a}_\in (k',\sigma')\epsilon^{(\sigma')\mu}(k')\;,
\end{equation}
and the orthogonality of the ``in''-and ``out''-states in conjunction with the fact that one-particle momentum eigenstates are stable, we obtain
\begin{eqnarray}
\prescript{}{\out}{\bigg\langle k';\sigma'\bigg|}\hat{V}^{\mu}(0)\ket{k^2+\frac{u^2}{2};\lambda'^2}_\in
&=&\sqrt{\hbar}(2\pi\hbar)^32k'^0\left[\epsilon^{(\lambda'_1)\mu}\left(k_1+\frac{u_1}{2}\right)\delta^{(3)}\left(\mathbf{k'}-\mathbf{k}_2-\frac{\mathbf{u}_2}{2}\right)\delta_{\sigma' \lambda_2'}+(1\leftrightarrow 2)\right]\nonumber\\
&&+ \tilde{\Delta}_R^{\mu\nu} \left(k_1+k_2+\frac{u_1+u_2}{2}-k'\right) \prescript{}{\out}{\bigg\langle k';\sigma'\bigg|}\hat{\rho}_\nu(0)\ket{k^2+\frac{u^2}{2};\lambda'^2}_\in\;. \label{sp_V}
\end{eqnarray}
Here we also used the Fourier decomposition of the propagator.
Inserting Eq. \eqref{sp_V} into Eq. \eqref{Phi_1}, we have
\begin{eqnarray}
&&\prescript{}{\in}{\bra{k^2-\frac{u^2}{2};\lambda^2}}\hat{\Phi}^{\mu\nu}(k)\ket{k^2+\frac{u^2}{2};\lambda'^2}_\in \nonumber\\
&=&-i  \sum_{\sigma'}\momint{k'}{}{}\bigg( \prescript{}{\out}{\bbra{k';\sigma'}}: \hat{\rho}^\nu (0): \ket{k^2+\frac{u^2}{2};\lambda'^2}_\in\nonumber\\
&&\times \bigg\{\sqrt{\hbar}(2\pi\hbar)^32k'^0\left[\epsilon^{*(\lambda_1)\mu}\left(k_1-\frac{u_1}{2}\right)\delta^{(3)}\left(\mathbf{k'}-\mathbf{k}_2+\frac{\mathbf{u}_2}{2}\right)\delta_{\sigma' \lambda_2}+(1\leftrightarrow 2)\right]\nonumber\\
&&+   \tilde{\Delta}_R^{*\mu\alpha} \left(k-\frac{u_1+u_2}{2}\right) \prescript{}{\in}{\bra{k^2-\frac{u^2}{2};\lambda^2}}:\hat{\rho}^{\dagger}_\alpha(0):\bket{k';\sigma'}_\out\bigg\} -\prescript{}{\in}{\bra{k^2-\frac{u^2}{2};\lambda^2}}:\hat{\rho}^{\dagger\mu}(0):\bket{k';\sigma'}_\out \nonumber\\
&&\times \bigg\{\sqrt{\hbar}(2\pi\hbar)^32k'^0\left[\epsilon^{(\lambda_1')\nu}\left(k_1+\frac{u_1}{2}\right)\delta^{(3)}\left(\mathbf{k'}-\mathbf{k}_2-\frac{\mathbf{u}_2}{2}\right)\delta_{\sigma' \lambda_2'}+(1\leftrightarrow 2)\right]\nonumber\\
&&+  \tilde{\Delta}_R^{\nu\alpha} \left(k+\frac{u_1+u_2}{2}\right) \prescript{}{\out}{\bbra{k';\sigma'}}:\hat{\rho}_\alpha(0):\ket{k^2+\frac{u^2}{2};\lambda'^2}_\in\bigg\}\bigg)\delta^{(4)}\left( k  +k' -k_1-k_2  \right) \;.\label{eq:phi_expectation}
\end{eqnarray}
Notice that we may rewrite the expectation value of a source term as
\begin{eqnarray}
&&\prescript{}{\out}{\bbra{k';\sigma'}} \hat{\rho}^\nu (0) \ket{k^2+\frac{u^2}{2};\lambda'^2}_\in \nonumber\\
&=&\sum_{\sigma=0}^3 g^{\sigma\sigma} \epsilon^{(\sigma)\nu}\left(k +\frac{u_1+u_2}{2}\right) \epsilon^{*(\sigma)\alpha}\left(k +\frac{u_1+u_2}{2}\right)\prescript{}{\out}{\bbra{k';\sigma'}} \hat{\rho}_\alpha (0) \ket{k^2+\frac{u^2}{2};\lambda'^2}_\in
\label{eq:rho_proj}\;,
\end{eqnarray}
where we introduced a timelike polarization vector $\epsilon^{(0)\mu}(k)\coloneqq k^\mu/k$, such that
\begin{equation}
    \sum_{\sigma'=0}^3 g^{\sigma'\sigma'} \epsilon^{(\sigma')\mu}(k)\epsilon^{*(\sigma')\nu}(k)=E^{\mu\nu}+K^{\mu\nu}=g^{\mu\nu}\;.
\end{equation}
Note that the term containing the timelike polarization vector in the sum in Eq. \eqref{eq:rho_proj} is actually of higher order in $\hbar$. To see this, consider the action of the respective four-momentum on the source term,
\begin{eqnarray}
    \left(k^\alpha+\frac{u_1^\alpha+u_2^\alpha}{2}\right)\prescript{}{\out}{\bbra{k';\sigma'}} \hat{\rho}_\alpha (0) \ket{k^2+\frac{u^2}{2};\lambda'^2}_\in &=& \prescript{}{\out}{\bbra{k';\sigma'}} \left[\hat{\rho}_\alpha (0),\hat{P}^\alpha\right] \ket{k^2+\frac{u^2}{2};\lambda'^2}_\in\nonumber\\
    &=& i\hbar \prescript{}{\out}{\bbra{k';\sigma'}} \left(\partial\cdot \hat{\rho}\right) (0) \ket{k^2+\frac{u^2}{2};\lambda'^2}_\in\;.\label{eq:timelike_higher_order}
\end{eqnarray}
If the source was conserved, this term would vanish identically. Even though we do not assume this, we will find that this term will not contribute to the Boltzmann equation to first order in $\hbar$.

The contraction of a polarization vector with a source term can be related to the transfer-matrix elements through \cite{DeGroot:1980dk}
\begin{eqnarray}
\epsilon^{*(\sigma)\alpha}\left(k\right)\prescript{}{\out}{\bra{k';\sigma'}} :\hat{\rho}_\alpha (0): \ket{k^2;\lambda'^2}_\in &=&-\frac{1}{\sqrt{\hbar}} \bra{k,k';\sigma,\sigma'} \hat{t}\ket{k^2;\lambda'^2}\;.\label{eq:source_to_scattering}
\end{eqnarray}
From this equation and Eq. \eqref{eq:timelike_higher_order} we deduce that all transfer-matrix elements in which one of the polarizations is timelike, i.e., where $\sigma=0$, are one order higher in $\hbar$ than their counterparts where all polarizations are spacelike.
Next we note that the retarded propagator of a massive vector boson may be written as
\begin{eqnarray}
\tilde{\Delta}_R^{\mu\nu}(k)&=&\hbar^2\left(\frac{E^{\mu\nu} }{m^2}- \frac{K^{\mu\nu}}{k^2-m^2+i\eta k^0}\right)\nonumber\\
&=&\tilde{\Delta}_R(k) \left[\sum_{\sigma=0}^3 g^{\sigma\sigma} \epsilon^{(\sigma)\mu}(k)\epsilon^{*(\sigma)\nu}(k)-\frac{k^2}{m^2}E^{\mu\nu}\right]\nonumber\\
&=&\tilde{\Delta}_R(k) \sum_{\sigma=0}^3 g^{\sigma\sigma} \epsilon^{(\sigma)\mu}(k)\epsilon^{*(\sigma)\nu}(k)\left(1-\frac{k^2}{m^2}\delta_{\sigma 0}\right)\;,\label{eq:propagator}
\end{eqnarray}
where we defined the scalar retarded propagator $\tilde{\Delta}_R(k)\coloneqq -\hbar^2 (k^2-m^2+i\eta k^0)^{-1}$ and used the fact that $\eta$ is an infinitesimal quantity prescribing which contour to take in the complex plane.

Inserting Eqs. \eqref{eq:rho_proj}, \eqref{eq:source_to_scattering} and \eqref{eq:propagator} into Eq. \eqref{eq:phi_expectation}, we obtain 
\begin{eqnarray}
&&\prescript{}{\in}{\bra{k^2-\frac{u^2}{2};\lambda^2}}\hat{\Phi}^{\mu\nu}(k)\ket{k^2+\frac{u^2}{2};\lambda'^2}_\in \nonumber\\
&=&-i\sum_{\sigma,\sigma'=0}^3g^{\sigma\sigma}g^{\sigma'\sigma'}\Bigg(\delta\left(k^0-k_1^0-k_2^0+\sqrt{\left(\mathbf{k}_2-\frac{\mathbf{u}_2}{2}\right)^2+m^2}\right) \delta^{(3)}\left(\mathbf{k}-\mathbf{k}_1-\frac{\mathbf{u}_2}{2}\right)\delta_{\sigma' \lambda_1} \nonumber\\
&&\times \bra{k+\frac{u_1+u_2}{2},k_2-\frac{u_2}{2};\sigma,\lambda_2} \hat{t} \ket{k^2+\frac{u^2}{2};\lambda'^2}  \epsilon^{(\sigma)\nu}\left(k+\frac{u_1+u_2}{2}\right)\epsilon^{*(\sigma')\mu}\left(k-\frac{u_1+u_2}{2}\right)\nonumber\\
&&+ (1\leftrightarrow 2)- \bigg[ \delta\left(k^0-k_1^0-k_2^0+\sqrt{\left(\mathbf{k}_2+\frac{\mathbf{u}_2}{2}\right)^2+m^2}\right) \delta^{(3)}\left(\mathbf{k}-\mathbf{k}_1+\frac{\mathbf{u}_2}{2}\right)\delta_{\sigma\lambda_1'}\nonumber\\
&&\times \bra{k^2-\frac{u^2}{2};\lambda^2}\hat{t}^{\dagger}\ket{k-\frac{u_1+u_2}{2},k_2+\frac{u_2}{2};\sigma',\lambda_2'}\epsilon^{*(\sigma')\mu}\left(k-\frac{u_1+u_2}{2}\right)\epsilon^{(\sigma)\nu}\left(k+\frac{u_1+u_2}{2}\right)\nonumber\\
&&+(1\leftrightarrow 2) \bigg] - \frac{1}{\hbar}\sum_{\sigma''} \momint{k'}{}{} \epsilon^{*(\sigma')\mu}\left(k-\frac{u_1+u_2}{2}\right) \epsilon^{(\sigma)\nu}\left(k+\frac{u_1+u_2}{2}\right)  \nonumber\\
&&\times  \bra{k^2-\frac{u^2}{2};\lambda^2}\hat{t}^{\dagger}\ket{k-\frac{u_1+u_2}{2},k';\sigma',\sigma''} \bra{k',k+\frac{u_1+u_2}{2};\sigma'',\sigma}\hat{t}\ket{k^2+\frac{u^2}{2};\lambda'^2}\nonumber\\
&&\times  \left\{ \tilde{\Delta}_R \left(k-\frac{u_1+u_2}{2}\right)\left[1-\frac{\left(k-\frac{u_1+u_2}{2}\right)^2}{m^2}\delta_{\sigma' 0}\right]-  \tilde{\Delta}_R^* \left(k+\frac{u_1+u_2}{2}\right)\left[1-\frac{\left(k+\frac{u_1+u_2}{2}\right)^2}{m^2}\delta_{\sigma 0}\right]  \right\}\nonumber\\
&&\times \delta^{(4)}\left( k  +k' -k_1-k_2  \right)\Bigg) \;,\label{eq:exp_phi_2}
\end{eqnarray}
where we inserted identities in order to be able to factor out the sums over $\sigma$ and $\sigma'$. 
In the second and fourth line of Eq. \eqref{eq:exp_phi_2} we separate the real and imaginary parts of $\hat{t}$ and $\hat{t}^\dagger$ in order to make use of the optical theorem
\begin{eqnarray}
\frac{i}{2}  \bra{k^2;\lambda^2} \hat{t}-\hat{t}^\dagger \ket{p^2;\lambda'^2}&=&-\frac{(2\pi\hbar)^4}{4} \sum_{\rho^2}\momint{q}{}{1}\momint{q}{}{2} \delta^{(4)}(q_1+q_2-k_1-k_2)\nonumber\\
&\times& \bra{k^2;\lambda^2}\hat{t}\ket{q^2;\rho^2}\bra{q^2;\rho^2}\hat{t}^\dagger\ket{p^2;\lambda'^2}\;.\label{optical_theorem}
\end{eqnarray}
In order to arrive at the expression in the main text, we furthermore need to rewrite the scattering-matrix elements as
\begin{equation}
\bra{k,k';\sigma,\sigma'} \hat{t}\ket{k^2;\lambda'^2}
=  \epsilon_\alpha^{*(\sigma)}(k)\epsilon_\beta^{*(\sigma')}(k') \epsilon_\gamma^{(\lambda_1')}(k_1) \epsilon_\delta^{(\lambda_2')}(k_2) M^{\alpha\beta\gamma\delta}(k,k',k_1,k_2)\;,\label{eq:t_to_M}
\end{equation}
where $M$ is the tree-level vertex function of the theory. In the remainder of this work we will assume that $M^*=M$, similar to Ref. \cite{Wagner:2022amr}. Note that the vertex fulfills $M^{\mu\nu\alpha\beta}=M^{\nu\mu\alpha\beta}=M^{\mu\nu\beta\alpha}=M^{\alpha\beta\mu\nu}$.

Truncating the sum in Eq. \eqref{kin_coll_exp} at $n=2$, inserting Eqs. \eqref{eq:exp_phi_2}--\eqref{eq:t_to_M} as well as using the completeness relation of the polarization vectors, we find for the components of the collision term that are orthogonal to the four-momentum 
\begin{equation}
    \mathcal{C}^{\mu\nu}_{\perp,\text{on-shell}}(x,k) =\mathcal{C}^{\mu\nu}_{\perp,\mathbf{k}\s}+\mathcal{C}^{\mu\nu}_{\perp,\s}\;.
\end{equation}
Here, $\mathcal{C}_{\perp,\k\s}^{\mu\nu}$ is defined as
\begin{subequations}\label{eq:coll_terms}
\begin{eqnarray}
\mathcal{C}^{\mu\nu}_{\perp,\k\s} &\coloneqq& \frac{(2\pi\hbar)^7}{4} \int \d^4 \bar{x}^2 \, \int \frac{\d^3 \mathbf{k}_1}{(2\pi\hbar)^32k_1^0}\int \frac{\d^3 \mathbf{k}_2}{(2\pi\hbar)^32k_2^0}\int \frac{\d^3 \mathbf{k}'}{(2\pi\hbar)^32k'^0}  \int \frac{\d^4 u^2}{(2\pi\hbar)^8}  e^{\frac{i}{\hbar}(u_1\cdot \bar{x}_1+u_2\cdot\bar{x}_2)}M^{\gamma_1\gamma_2 \delta_1\delta_2}M^{\zeta_1\zeta_2 \eta_1\eta_2}\nonumber\\
&\times&K^\mu_{\mu'}K^\nu_{\nu'}\left(K-\frac{U_1+U_2}{2}\right)^{\mu'\alpha}\left(K+\frac{U_1+U_2}{2}\right)^{\nu'\beta}  \bigg[ W_{\text{on-shell}}^{\alpha_1\beta_1}(x+\bar{x}_1,k_1)W_{\text{on-shell}}^{\alpha_2\beta_2}(x+\bar{x}_2,k_2) \nonumber\\ 
&&\times  g_{\alpha\delta_1}g_{\beta\zeta_1}\left(K_1-\frac{U_1}{2}\right)_{\gamma_1\alpha_1}\left(K_2-\frac{U_2}{2}\right)_{\gamma_2\alpha_2}K'_{\delta_2\zeta_2}\left(K_1+\frac{U_1}{2}\right)_{\beta_1\eta_1}\left(K_2+\frac{U_2}{2}\right)_{\beta_2\eta_2}\delta^{(4)}(k+k'-k_1-k_2) \nonumber\\
&& - \frac12 W_{\text{on-shell}}^{\alpha_1\beta_1}\left(x+\bar{x}_1,k-\frac{u_2}{2}\right)W_{\text{on-shell}}^{\alpha_2\beta_2}(x+\bar{x}_2,k') g_{\alpha\alpha_1} g_{\beta\gamma_1}  \left(K'-\frac{U_2}{2}\right)_{\gamma_2\alpha_2}K_{1,\delta_1\zeta_1}K_{2,\delta_2\zeta_2}\nonumber\\
&&\times
\left(K+\frac{U_1-U_2}{2}\right)_{\beta_1\eta_1}\left(K'+\frac{U_2}{2}\right)_{\beta_2\eta_2}\delta^{(4)}\left(k+k'-k_1-k_2+\frac{u_1}{2}\right)\nonumber\\
&& - \frac12 W_{\text{on-shell}}^{\alpha_1\beta_1}\left(x+\bar{x}_1,k+\frac{u_2}{2}\right)W_{\text{on-shell}}^{\alpha_2\beta_2}(x+\bar{x}_2,k') g_{\alpha\eta_1} g_{\beta\beta_1}  \left(K'-\frac{U_2}{2}\right)_{\gamma_2\alpha_2}K_{1,\delta_1\zeta_1}K_{2,\delta_2\zeta_2}\nonumber\\
&&\times 
\left(K+\frac{U_2-U_1}{2}\right)_{\gamma_1\alpha_1}\left(K'+\frac{U_2}{2}\right)_{\beta_2\eta_2}\delta^{(4)}\left(k+k'-k_1-k_2-\frac{u_1}{2}\right)\bigg]\;.\label{eq:C1}
\end{eqnarray}
This term will turn out to describe collisions exchanging both momentum and spin, cf. the discussion in the main text.
The term $\mathcal{C}_{\perp,\s}^{\mu\nu}$ will be responsible for the collisions exchanging only spin, and reads
\begin{eqnarray}
\mathcal{C}^{\mu\nu}_{\perp,\s} &\coloneqq& \frac{i(2\pi\hbar)^3}{4} \int \d^4 \bar{x}^2 \, \int \frac{\d^3 \mathbf{k}_2}{(2\pi\hbar)^32k_2^0} \int \frac{\d^4 u^2}{(2\pi\hbar)^8}  e^{\frac{i}{\hbar}(u_1\cdot \bar{x}_1+u_2\cdot\bar{x}_2)}M^{\gamma_1\gamma_2 \delta_1\delta_2}\nonumber\\
&&\times \left(K_2-\frac{U_2}{2}\right)_{\gamma_2\alpha_2}\left(K_2+\frac{U_2}{2}\right)_{\beta_2\delta_2} \left(K-\frac{U_1+U_2}{2}\right)^{\mu'\alpha}\left(K+\frac{U_1+U_2}{2}\right)^{\nu'\beta} K^\mu_{\mu'}K^\nu_{\nu'} \nonumber\\
&&\times\bigg[ W_{\text{on-shell}}^{\alpha_1\beta_1}\left(x+\bar{x}_1,k-\frac{u_2}{2}\right)W_{\text{on-shell}}^{\alpha_2\beta_2}(x+\bar{x}_2,k_2)g_{\alpha\alpha_1} g_{\beta\delta_1} 
\left(K+\frac{U_1-U_2}{2}\right)_{\beta_1\delta_1}\nonumber\\
&& \;\;-  W_{\text{on-shell}}^{\alpha_1\beta_1}\left(x+\bar{x}_1,k+\frac{u_2}{2}\right)W_{\text{on-shell}}^{\alpha_2\beta_2}(x+\bar{x}_2,k_2) g_{\alpha\gamma_1}   g_{\beta\beta_1}
\left(K+\frac{U_2-U_1}{2}\right)_{\gamma_1\alpha_1}\bigg]\;.\label{eq:C2}
\end{eqnarray}
\end{subequations}
Note that, as mentioned in Ref. \cite{Wagner:2022amr}, it can be shown that this term gives a correction to the drift term and a Vlasov-like contribution on the left-hand side of the Boltzmann equation.

In these expressions, we already used that the Wigner functions entering the Boltzmann equation are on the mass-shell, cf. App. \ref{app:offshell}. Furthermore, we employed the relation
\begin{equation}
    \tilde{\Delta}_R\left(k\right)-\tilde{\Delta}^*_R (k)=2\pi i\hbar^2 \delta(k^2-m^2)\;.
\end{equation}
In Eqs. \eqref{eq:coll_terms}, it can be seen that to first order the terms in the sum in Eq. \eqref{eq:exp_phi_2} where $\sigma=0$ or $\sigma'=0$ do not contribute to $\mathcal{C}_{\perp,\text{on-shell}}^{\mu\nu}$. This is due to the fact that the transfer-matrix elements containing timelike polarizations are one order higher in the $\hbar$-gradient expansion than their counterparts containing only spacelike polarization, cf. Eq. \eqref{eq:timelike_higher_order}.

In order to arrive at Eq. \eqref{eq:coll_int_full} in the main text, we neglect the pure-spin exchange term $\mathcal{C}^{\mu\nu}_{\perp,\s}$, approximate (for $j\inmath\{1,2\}$) 
\begin{equation}
W^{\mu\nu}(x+\bar{x}_j,k)\simeq W^{\mu\nu}(x,k)+\bar{x}_j\cdot \partial W^{\mu\nu}(x,k)
\end{equation} 
and perform the $\d^4 \bar{x}^2$-integrations.

\section{Proof that the off-shell terms cancel}
\label{app:offshell}

In this appendix, we show that the off-shell terms in the kinetic equation vanish to first order in $\hbar$.
Acting with the Bopp operators on the collision integral \eqref{eq:coll_int_1} and neglecting terms of higher order in $\hbar$, it holds that
\begin{equation}
\left(D^*\cdot D^*-m^2\right)C^{\mu\nu}_\perp(x,k)=\hbar \frac{2\hbar^2}{(2\pi\hbar)^4}K^\mu_\alpha K^\nu_\beta  \int \d^4 v e^{-ik\cdot v/\hbar} \avg{\hat{\rho}^{\dagger\alpha}(x+v/2)\hat{\rho}^\beta(x-v/2)} =: \hbar Z^{\mu\nu}(x,k)\;.
\end{equation}
Note that it is sufficient to consider the components of $C^{\mu\nu}$ orthogonal to $k^\mu$ since only these enter into Eq. \eqref{Boltzmann_f}.
Since $Z^{*\nu\mu}=Z^{\mu\nu}$, we have that
\begin{equation}
\Re Z^{\mu\nu}=Z_S^{\mu\nu}\;,\quad \Im Z^{\mu\nu}=-iZ_A^{\mu\nu}\;.
\end{equation}
Splitting the equation of motion for $C^{\mu\nu}$ into real and imaginary parts, we obtain
\begin{align}
(k^2-m^2)\Re C_{\perp,S}^{\mu\nu}+\hbar k\cdot \partial \Im C^{\mu\nu}_{\perp,S}&= \hbar Z_S^{\mu\nu}\;,&\quad (k^2-m^2)\Re C_{\perp,A}^{\mu\nu}+\hbar k\cdot \partial \Im C^{\mu\nu}_{\perp,A}&=0\;,\nonumber\\
(k^2-m^2)\Im C_{\perp,A}^{\mu\nu}-\hbar k\cdot \partial \Re C^{\mu\nu}_{\perp,A} &= -i\hbar Z_A^{\mu\nu}\;,&\quad (k^2-m^2)\Im C_{\perp,S}^{\mu\nu}-\hbar k\cdot \partial \Re C^{\mu\nu}_{\perp,S}&=0\;.
\end{align}
Subtracting the fourth equation from the second one multiplied by $i$, we have
\begin{equation}
(k^2-m^2)\left( i\Re C_{\perp,A}^{\mu\nu}-\Im C_{\perp,S}^{\mu\nu}  \right) +\hbar k\cdot \partial \left( \Re C_{\perp,S}^{\mu\nu} +i\Im C_{\perp,A}^{\mu\nu}\right)=0\;.
\end{equation}
Making use of Eqs. \eqref{C_M_def}, this becomes
\begin{equation}
(k^2-m^2)\C^{\mu\nu}_\perp=\hbar k\cdot \partial \delta M^{\mu\nu}_\perp\;,\label{equality_off_shell}
\end{equation}
which implies that the collision kernel may be expanded as
\begin{equation}
\C^{\mu\nu}_\perp=\delta(k^2-m^2)\left(\C^{(0),\mu\nu}_\perp+\C^{(1),\mu\nu}_{\perp,\text{on-shell}}  \right)+\frac{\hbar}{k^2-m^2} k\cdot \partial \delta M^{\mu\nu}_\perp+\mathcal{O}(\hbar^2)\;.\label{expansion_C}
\end{equation}
Remembering the mass-shell equations \eqref{massshell_eqs}
we get to first order in $\hbar$
\begin{equation}
W^{\mu\nu}_\perp(x,k)=\delta(k^2-m^2)\left[W^{(0),\mu\nu}_\perp(x,k)+\hbar W^{(1),\mu\nu}_{\perp,\text{on-shell}}(x,k)\right]+\frac{\hbar}{k^2-m^2} \delta M^{\mu\nu}_\perp\;. 
\end{equation}
Inserting this solution into the kinetic equation $k\cdot \partial W^{\mu\nu}_\perp(x,k)=\C^{\mu\nu}_\perp$ and making use of Eq. \eqref{expansion_C}, we obtain
\begin{eqnarray}
&&\delta(k^2-m^2) k\cdot \partial\left[W^{(0),\mu\nu}_\perp(x,k)+ \hbar W^{(1),\mu\nu}_{\perp,\text{on-shell}}(x,k)\right]+\frac{\hbar}{k^2-m^2}k\cdot \partial \delta M^{\mu\nu}_\perp \nonumber\\
&=&\delta(k^2-m^2)\left(\C^{(0),\mu\nu}_\perp+\C^{(1),\mu\nu}_{\perp,\text{on-shell}}  \right)+\frac{\hbar}{k^2-m^2} k\cdot \partial \delta M^{\mu\nu}_\perp \;.
\end{eqnarray}
It is straightforward to see that the off-shell terms cancel and the Boltzmann equation is on shell:
\begin{equation}
 k\cdot \partial W^{\mu\nu}_{\perp,\text{on-shell}}(x,k)=\C^{\mu\nu}_{\perp,\text{on-shell}}  \;.
\end{equation}

\bibliography{biblio_paper_long}

\end{document}